\documentclass[sigconf]{acmart}

\usepackage{graphicx}
\usepackage{amsmath}

\usepackage{amssymb}
\usepackage{booktabs}
\usepackage{balance}
\usepackage{multirow}
\usepackage{mathrsfs}
\usepackage{subcaption}
\usepackage{bbding}
\usepackage{pifont}
\usepackage[ruled]{algorithm2e} % For algorithms

\AtBeginDocument{%
  \providecommand\BibTeX{{%
    \normalfont B\kern-0.5em{\scshape i\kern-0.25em b}\kern-0.8em\TeX}}}
\copyrightyear{2023}
\acmYear{2023}
\setcopyright{rightsretained}
\acmConference[MM '23] {Proceedings of the 31st ACM International Conference on Multimedia}{October 29--November 3, 2023}{Ottawa, ON, Canada.}
\acmBooktitle{Proceedings of the 31st ACM International Conference on Multimedia (MM '23), October 29--November 3, 2023, Ottawa, ON, Canada}
\acmPrice{}
\acmISBN{979-8-4007-0108-5/23/10}
\acmDOI{10.1145/3581783.3612463}

\makeatletter
\gdef\@copyrightpermission{
  \begin{minipage}{0.3\columnwidth}
   \href{https://creativecommons.org/licenses/by/4.0/}{\includegraphics[width=0.90\textwidth]{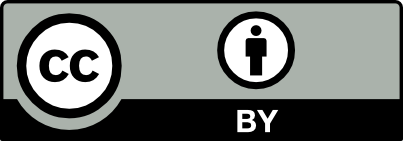}}
  \end{minipage}\hfill
  \begin{minipage}{0.7\columnwidth}
   \href{https://creativecommons.org/licenses/by/4.0/}{This work is licensed under a Creative Commons Attribution International 4.0 License.}
  \end{minipage}
  \vspace{5pt}
}
\makeatother

\begin{document}

\title{Towards Balanced Active Learning for Multimodal Classification}

\author{Meng Shen}
\affiliation{%
  \institution{Nanyang Technological University}
  \country{Singapore}}
\email{meng005@e.ntu.edu.sg}

\author{Yizheng Huang}
\affiliation{%
  \institution{Institute for Infocomm Research A*STAR}
  \country{Singapore}}
\email{huangyz0918@ieee.org}

\author{Jianxiong Yin}
\affiliation{%
  \institution{NVIDIA AI Tech Center}
  \country{Singapore}}
\email{jianxiongy@nvidia.com}

\author{Heqing Zou}
\affiliation{%
  \institution{Nanyang Technological University}
  \country{Singapore}}
\email{heqing001@e.ntu.edu.sg}

\author{Deepu Rajan}
\affiliation{%
  \institution{Nanyang Technological University}
  \country{Singapore}}
\email{asdrajan@ntu.edu.sg}

\author{Simon See}
\affiliation{%
  \institution{NVIDIA AI Tech Center}
  \country{Singapore}}
\email{ssee@nvidia.com}

\renewcommand{\shortauthors}{Meng Shen et al.} 

%%%%%%%%% ABSTRACT
\begin{abstract}
Training multimodal networks requires a vast amount of data due to their larger parameter space compared to unimodal networks. Active learning is a widely used technique for reducing data annotation costs by selecting only those samples that could contribute to improving model performance. However, current active learning strategies are mostly designed for unimodal tasks, and when applied to multimodal data, they often result in biased sample selection from the dominant modality. This unfairness hinders balanced multimodal learning, which is crucial for achieving optimal performance. To address this issue, we propose three guidelines for designing a more balanced multimodal active learning strategy. Following these guidelines, a novel approach is proposed to achieve more fair data selection by modulating the gradient embedding with the dominance degree among modalities. Our studies demonstrate that the proposed method achieves more balanced multimodal learning by avoiding greedy sample selection from the dominant modality. Our approach outperforms existing active learning strategies on a variety of multimodal classification tasks. Overall, our work highlights the importance of balancing sample selection in multimodal active learning and provides a practical solution for achieving more balanced active learning for multimodal classification.
\end{abstract}

\begin{CCSXML}
<ccs2012>
   <concept>
       <concept_id>10010147.10010257.10010282.10011304</concept_id>
       <concept_desc>Computing methodologies~Active learning settings</concept_desc>
       <concept_significance>500</concept_significance>
       </concept>
 </ccs2012>
\end{CCSXML}

\ccsdesc[500]{Computing methodologies~Active learning settings}

\keywords{active learning, multimodal learning}



\maketitle

%%%%%%%%% BODY TEXT
\section{Introduction}
\label{sec:intro}

Multimodal classification, as one of the classical multimodal learning tasks, aims to exploit complementary information inherent in multimodal data to achieve better classification performance. To this end, deep learning strategies have been implemented to train large-scale multimodal deep neural networks \cite{DBLP:journals/pami/BaltrusaitisAM19, DBLP:journals/access/GuoWW19}. However, such networks require an enormous amount of data to learn from, given their huge number of parameters. To reduce data cost, active learning (AL) is used to select a subset of more informative and distinctive unlabeled data samples for label assignment by oracles. Consequently, large networks can maintain performance while utilizing a smaller labeling budget. Most existing active learning algorithms are designed for unimodal tasks such as image classification \cite{Coreset-SenerS18, DBLP:conf/cvpr/BeluchGNK18}, object detection \cite{Influence-iccv/LiuDZLDH21, DBLP:conf/cvpr/YuanWFLXJY21} and language modeling \cite{Cold-Start-AL/emnlp/YuanLB20, emnlp/margatina2021active}. The objective is to select samples that have high uncertainty in them, carry novel knowledge for model training and those with distinctive features. However, there has been significantly less research reported on the design of effective active learning strategies for multimodal learning \cite{AL-survey}.

In this paper, we initially examine the performance of existing active learning strategies in selecting multimodal data. Our experiments reveal that these strategies tend to focus more on the dominant modality rather than fairly considering all modalities. For instance, in an image-text classification task, if the text contributes more to model optimization, active learning strategies may exhibit a bias towards the more distinguishable text modality by selecting valuable text samples and disregarding the informativeness of image samples. As a result, the selected multimodal dataset could become unbalanced, with insufficient information from the image modality, potentially leading to a degraded image model backbone. Recent works \cite{balance_mm/DBLP:conf/cvpr/PengWD0H22, DBLP:conf/cvpr/WangTF20, DBLP:conf/icml/HuangLZYH22, DBLP:conf/icml/WuJCG22} point out that balancing the training and optimization of all modalities is a key factor for successful multimodal learning. Similarly, it is crucial to design active learning strategies that can select multimodal data with fairness among all modalities to assist balanced multimodal learning.

Based on our findings, we develop a \textbf{B}alanced \textbf{M}ulti\textbf{m}odal \textbf{A}ctive \textbf{L}earning (\textbf{BMMAL}) algorithm that selects multimodal data by fairly considering each modality present in the data. In our approach, we choose the gradient embedding of model parameters, as it reflects the impact on model training and captures the diversity of data samples. However, we examine how the previous gradient embedding method \cite{badge-iclr/AshZK0A20} fails to select balanced multimodal data. To ensure fairness, we individually assess the contribution of each modality feature by examining the Shapley value, which attributes its contribution to the final multimodal prediction. We then apply modulation on the gradient embedding to penalize samples with dominant modalities. Lastly, a clustering seed initialization algorithm is employed to select diverse multimodal data with a significant influence on model training.

In summary, our main contributions are as follows:

\begin{itemize}
    \item We empirically show that most existing active learning strategies fail to select a balanced multimodal dataset. We analyze how to improve the current gradient embedding based active learning strategy to rectify this.
    \item We propose a method to modulate the gradient embedding on sample-level to select more balanced multimodal candidates.
    \item We conduct experiments on three multimodal datasets to show that our proposed method treats multimodal data more equally and achieves better performance. 
\end{itemize}

%%%%%%%%% RELATED WORKS
\section{Related Works}
\label{sec.2}

\subsection{Active Learning}

Uncertainty-aware strategies attempt to utilize the data uncertainty or the model uncertainty as a criterion to locate unlabeled data points that the current model has less confidence about. One strategy is to utilize the posterior classification probability distribution by measuring its entropy \cite{DBLP:series/synthesis/2012Settles, DBLP:conf/icmcs/WangKTCP15}, or the margin between the most confident class and the second most confident class \cite{DBLP:conf/ecml/RothS06}. In addition, uncertainty can be evaluated as the variance of predictions generated by an ensemble of models \cite{DBLP:conf/cvpr/BeluchGNK18} or by multiple inferences with Monte-Carlo dropout as an alternative Bayesian approximation for static networks \cite{DBLP:conf/icml/GalG16}. Moreover, ALFA-Mix \cite{DBLP:conf/cvpr/ParvanehATHHS22} evaluates unlabeled samples by mixing their features with labeled samples and observing whether there is inconsistency among predictions from mixed features. DFAL \cite{Adversarial-deepfool} incorporates adversarial attack techniques \cite{DBLP:conf/cvpr/Moosavi-Dezfooli16} to select unlabeled data samples located close to the classification boundaries.

Diversity-aware strategies tend to select unlabeled data points whose features are as diverse as possible to minimize data redundancy. \cite{Nguyen2004ActiveLU} utilizes K-medoid algorithm \cite{kaufman1990finding} to select representative data centroids that minimize the total distance from other data samples to the nearest centroids. CoreSet \cite{Coreset-SenerS18} greedily selects unlabeled data samples that have maximum distances from their nearest neighbors. \cite{DBLP:journals/corr/abs-1906-07975} adopts the determinantal point process (DPP) to evaluate the diversity by calculating the determinant of the similarity matrix. Diversity-aware strategies can also be considered in the context of distribution matching, which aims to reduce the gap between the distributions of labeled and unlabeled samples in latent space or feature space. VAAL \cite{DBLP:conf/iccv/SinhaED19} trains a variational auto-encoder to construct the latent distribution of labeled samples and an adversarial network to distinguish labeled samples and unlabeled samples in the latent space. Moreover, the maximum mean discrepancy (MMD) \cite{DBLP:journals/ml/VieringKL19}, the $\mathcal{H}$-divergence \cite{DBLP:conf/wacv/SuTSLMC20} and the Wasserstein distance \cite{DBLP:conf/aistats/ShuiZGW20} are used to measure the distribution gap.

To achieve a better trade-off between informativeness and diversity, hybrid methods are developed with an awareness of both. Since diversity-aware strategies are orthogonal to most of uncertainty-aware strategies \cite{DBLP:conf/icml/HacohenDW22}, they could be easily combined together. ALFA-Mix \cite{DBLP:conf/cvpr/ParvanehATHHS22} adopts K-means clustering to further filter out samples to enhance diversity. BADGE \cite{badge-iclr/AshZK0A20} represents unlabeled data samples via gradient embedding of parameters of the last classifier layer and applies K-means++ \cite{DBLP:conf/soda/ArthurV07} to form a diverse data selection which still carries high uncertainty.

\subsection{Balanced Multimodal Learning}

Our work considers joint multimodal learning for classifications. Here, it has been found that the best unimodal networks could potentially outperform multimodal networks regardless of fusion mechanisms or regularization methods \cite{DBLP:conf/cvpr/WangTF20}. Recent works show that the degradation of multimodal learning could be due to unbalanced optimization among different modalities. In \cite{DBLP:conf/icml/HuangLZYH22}, the failure of multimodal learning is attributed to modality competition where only dominant modalities are fully explored by joint training. Similarly, \cite{DBLP:conf/icml/WuJCG22} demonstrates that multimodal learning greedily optimizes the dominant modalities and chooses to balance their training speeds. \cite{DBLP:conf/cvpr/WangTF20} propose to blend gradients with weights that are disproportional to the overfitting and generalization ratio of each modality so that each modality could be optimized in a balanced manner. \cite{balance_mm/DBLP:conf/cvpr/PengWD0H22} finds that fusion mechanisms such as concatenation and summation encourage the dominant modality to learn faster and thus develops gradient modulation to adaptively balance the training speed of each modality.

\begin{figure}[t]
  \centering
   \includegraphics[width=0.75\linewidth]{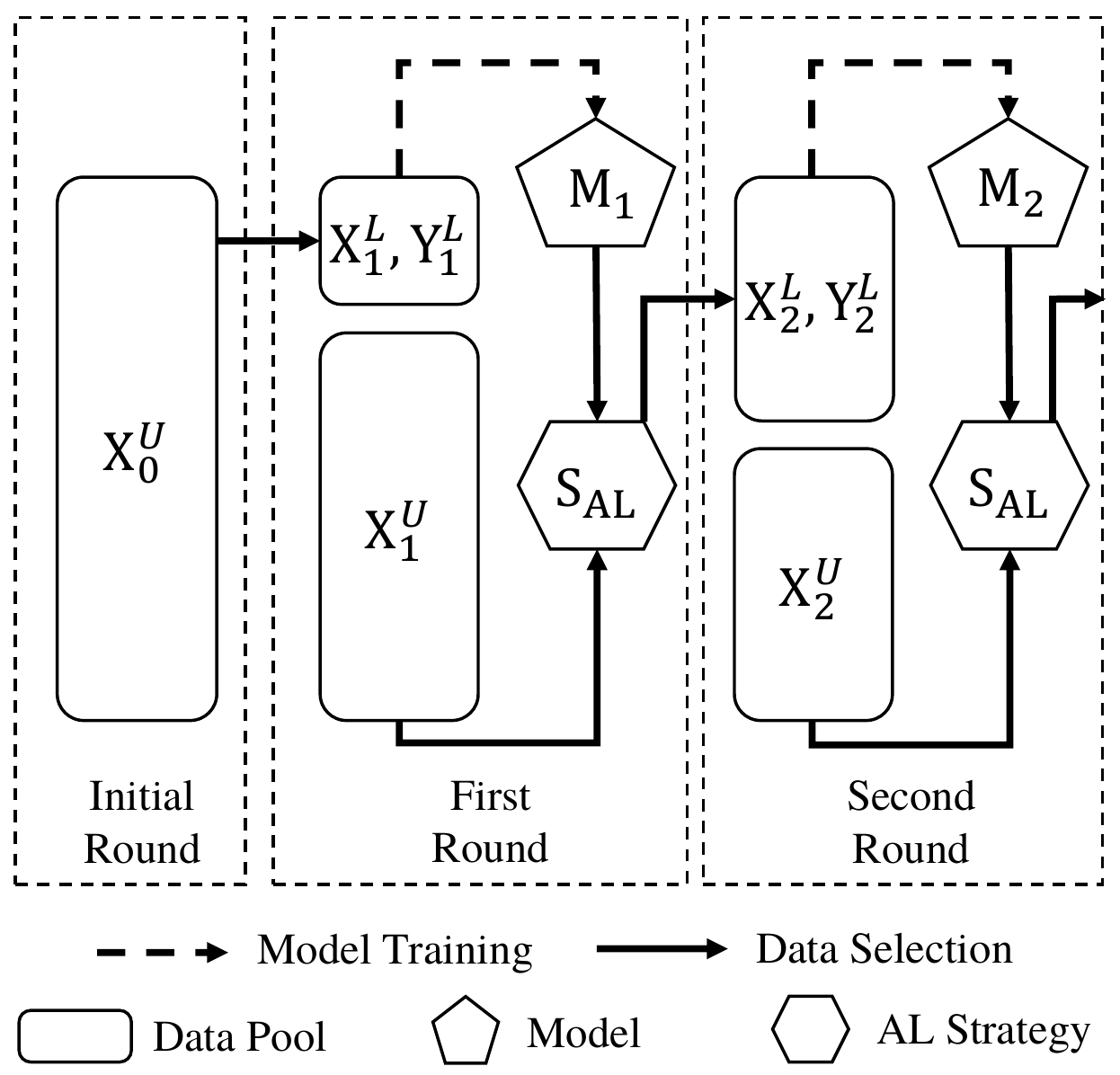}
   \caption{General active learning process. The dashed lines represent model training. The solid lines represent data selection.}
   \label{fig:active_learning}
\end{figure}

\section{Methodology}
\label{sec.3}

\subsection{Multimodal Active Learning Framework}

The general active learning process is shown in \textbf{Figure \ref{fig:active_learning}}. Initially, we are given a large unlabeled data pool $X_0^U = \{(x_{m_1}, \dots, x_{m_M})_{1 \dots n}\}$ of $n$ input data  with $M$ modalities and an empty labeled data pool $X_0^L = \emptyset$. The labeling budget of each round is set to $B$. In the first round of active learning, since there is no trained model to evaluate with, a subset $X_1^L$ containing $B$ multimodal data is randomly selected from $X_0^U$, and they will be assigned with true labels $Y_1^L$. After data selection, the unlabeled dataset becomes $X_1^U = X_0^U \setminus X_1^L$. The training dataset for the first round of model training consists of $X_1^L$ and $Y_1^L$. Starting from the second round, an active learning strategy $S_{AL}$ evaluates the trained model and unlabeled data in the last round using an acquisition function and selects a batch of candidates for label assignment to construct a new training dataset for the current round of model training. The processes of data selection and model training continue until the total labeling budget is run out or the target performance of the trained model is reached.

We then introduce our multimodal learning framework for classification task. $x_{m_1}$ and $x_{m_2}$ represent the input data from two different modalities. They are processed through encoders $\varphi_{m_1}$ and $\varphi_{m_2}$ respectively to extract unimodal features $z_{m_1} \in \mathbb{R}^{D_{m1}}$ and $z_{m_2} \in \mathbb{R}^{D_{m_2}}$.  We adopt concatenation, a wildly used late-fusion mechanism, to construct multimodal features $z_{mm} = z_{m_1} \oplus z_{m_2}$.\footnote{Other fusion mechanisms such as summation and NL-gate are implemented in our further experiments.} The unimodal and multimodal features are fed to unimodal classifiers $C_{m_1}$, $C_{m_2}$ and multimodal classifier $C_{mm}$ respectively to produce logits $f_{m_1}$, $f_{m_2}$ and $f_{mm}$ for classification. The final loss is the average cross-entropy loss $\mathcal{L}_{CE}$ of unimodal and multimodal logits with true labels $y$:   

\begin{equation}
    \mathcal{L}_{final} =  \frac{1}{3} [\mathcal{L}_{CE}(f_{m_1}, y)+\mathcal{L}_{CE}(f_{m_2}, y)+\mathcal{L}_{CE}(f_{mm}, y)].
\end{equation}

Once the model is trained, the unlabeled data samples are evaluated using an acquisition function and filtered for labeling.

\subsection{Analysis of Imbalance in AL}

We introduce one of the state-of-the-art active learning algorithm BADGE \cite{badge-iclr/AshZK0A20} and provide analysis of its imbalanced data selection over multimodal data samples. BADGE was the first to propose the replacement of features for embedding with the gradient of the weight of the last FC layer, which acts as the classifier. In our case, the last FC layer for multimodal classification is the multimodal classifier $C_{mm}$. The weight of classifier $W$ is a 2-dimensional matrix of size  $K \times D_{mm}$, where $K$ is the number of classes and $D_{mm}$ is the dimension of concatenated multimodal feature $D_{m_1} + D_{m_2}$. The corresponding multimodal cross-entropy loss can be expanded as
\begin{equation}
\begin{aligned}
    \mathcal{L}_{mm} & = - \sum_{i=1}^{K}y_i \cdot log \sigma (f_{mm})_i \\
    & = - \sum_{i=1}^{K}y_i \cdot log \frac{e^{z_{mm} \cdot W_i^T }}{\sum_{i=1}^K e^{z_{mm} \cdot W_i^T }},
\end{aligned}
\end{equation}
where $\sigma$ is softmax function and ${z_{mm} \cdot W_i^T}$ is the $i^{th}$ element of logits $f_{mm}$. The gradient embedding is defined as $g = \frac{\partial \mathcal{L}_{mm}}{\partial W}$, and it is a 2-D matrix of size $K \times D_{mm}$ where the $i^{th}$ row is
\begin{equation}
    g_i = (f_i - 1_{\hat{y}_{mm} = i}) z_{mm},
\label{eq:3}
\end{equation}
where $\hat{y}_{mm} = \underset{i\in[K]}{\mathrm{argmax}} [({f_{mm}})_i] $ is the pseudo label for unlabeled data samples. The gradient embedding is flattened into a vector for sampling. It not only carries the uncertainty of classification from the margin between logits $f_i$ and pseudo labels $\hat{y}_{mm}$, but also is representative enough due to the information present in $z_{mm}$.

However, in multimodal learning settings, identifying the source of uncertainty can be challenging. Upon examining the calculation of multimodal logits $f_i={z_{mm} \cdot W_i^T} = z_{m_1} \cdot ({W_i})_{m_1}^T +  z_{m_2} \cdot ({W_i})_{m_2}^T$, where $W_i$ is divided into two matrices $({W_i})_{m_1}$ and $({W_i})_{m_2}$, it is difficult to determine which modality carries more uncertainty and which carries less. To illustrate, for a visual event such as drawing, the visual modality contains more information and contributes more to multimodal logits by generating a larger output. The multimodal uncertainty calculation is thus skewing the visual uncertainty instead of considering both visual and auditory uncertainties fairly. From \textbf{Section \ref{sec.4.4}}, we find that BADGE does pay more attention to the dominant modality, which might potentially damage the performance of joint multimodal learning. Another limitation of BADGE is its inability to distinguish modality contributions. For instance, given two data samples with identical logits, we should prioritize the one with a more balanced contribution during data selection to facilitate balanced multimodal learning. However, the current BADGE algorithm cannot achieve this. Similarly, most conventional active learning algorithms lack this capability.

Hence, we develop a balanced multimodal active learning method that could avoid biased data selection towards the dominant modality to mitigate modality competition and assure that the trained multimodal network would not easily degenerate to the dominant modality. While our designed method is encouraged to pay more attention to the weaker modality, it is essential to ensure that it does not overly lean towards the weaker modality, as this may also harm the multimodal classification performance.

\subsection{Guidelines to Design Balanced MMAL}

To make existing AL strategies more suitable for balanced multimodal learning, it is necessary to inspect the individual modality contribution and reduce the contribution gap among different modalities. We empirically propose three guidelines for designing active learning strategies that treat each modality more equally. Let $\Phi_{m_i}(x)$ represent the contribution of the $i^{th}$ modality of data sample $x$ to the final model outcome , which should satisfy:

\begin{equation}
    \sum_{i=1}^{M}\Phi_{m_i}(x) = 1.
\end{equation}

We introduce the dominance degree $\rho(x)$ to quantify how severely  a data sample $x$ is dominated by the strongest modality:

\begin{equation}
    \rho(x) = \sum_{i=1}^{M}[max(\Phi_{m_1}(x),...,\Phi_{m_M}(x)) - \Phi_{m_i}(x)].
\end{equation}

We further partition the entire unlabeled dataset into multiple subsets $X=\{X_1, ... ,X_M\}$ for the ease of discussion. In each subset $X_i$, modality $m_i$ contributes the most:

\begin{equation}
\label{eq:6}
    \Phi_{m_i}(x) \ge \Phi_{m_j}(x), i \neq j, \forall x \in X_i.
\end{equation}

\textbf{Guideline 1}: For two multimodal data samples $x_{i}$ and $x_{j}$, if their acquisition scores of conventional active learning (CAL) strategies are equal, the one with more balanced unimodal contributions should have higher acquisition scores of balanced multimodal active learning strategies,

\begin{equation}
\begin{aligned}
    & a_{BMMAL}(x_{i}, \rho_i) > a_{BMMAL}(x_{j}, \rho_j), \rho_i < \rho_j, \\
    & \text{where }  a_{CAL}(x_{i}) = a_{CAL}(x_{j}), i \neq j.
\end{aligned}
\end{equation}

By following Guideline 1, data samples with more equal unimodal contributions are more likely to be selected. However, this does not guarantee that the stronger modality will be suppressed, nor does it ensure that the weaker modality will not be overly encouraged. Therefore, we introduce two additional guidelines.

\textbf{Guideline 2}: To avoid biased data selection favoring the stronger modality, the gap between the average acquisition scores of data samples dominated by the stronger modality and those dominated by the weaker modality should be reduced. In a two-modality case, where $m_1$ is the weaker modality and $m_2$ is the stronger modality (i.e. the average contribution of $m_1$ over the entire dataset is less than that of $m_2$, $\frac{1}{|X|}\sum_{x\in X}\Phi_{m_1}(x)  < \frac{1}{|X|}\sum_{x\in X}\Phi_{m_2}(x)$), we have

\begin{equation}
    \frac{\frac{1}{|X_1|}\sum_{x \subseteq X_1}a_{CAL}(x)}{\frac{1}{|X_2|}\sum_{x \subseteq X_2}a_{CAL}(x)} < \frac{\frac{1}{|X_1|}\sum_{x \subseteq X_1}a_{BMMAL}(x)}{\frac{1}{|X_2|}\sum_{x \subseteq X_2}a_{BMMAL}(x)}.
\end{equation}

\textbf{Guideline 3}: Lastly, to prevent biased data selection towards the weaker modality, it is necessary to ensure that the contribution of each modality to the acquisition score function $a_{BMMAL}$ is still proportional to its modality contribution to the model outcome on the sample-level. It ensures that the data samples are selected in a way that fairly represents the contributions of each modality to the actual model outcome.

In summary, Guideline 1 prioritizes the samples with more equal unimodal contributions. Guideline 2 and 3 work together to punish the stronger modality on the dataset-level but maintain the relationship between strong and weak modality on the sample-level, avoiding biases towards either the stronger or weaker modalities.

\subsection{Estimate Modality Contribution}

We show how we compute modality contribution $\Phi$. In the context of multimodal classification, balanced active learning should select data samples that fairly contribute to the performance of all modalities. To achieve this, it is essential to estimate the degree to which each modality of a given data sample contributes to the final multimodal prediction. 
One approach involves assessing modality importance by computing the disparity in model performance before and after the incorporation of a particular modality. Researchers have proposed various techniques to remove the information of one modality, such as masking \cite{frank-etal-2021-vision}, permutation \cite{PerceptualScore}, and empirical multimodally-additive projection (EMAP) \cite{EMAP/EMNLP}. Nonetheless, these attribution methods are ill-suited for active learning as they require ground truth labels to calculate model performance metrics, such as accuracy. As a result, these methods cannot be employed for estimating modality contribution for unlabeled data due to the absence of ground truth labels.

Therefore, we choose to use the Shapley value to estimate modality contribution without the need for true labels. The Shapley value \cite{shapley1953value} was proposed to fairly attribute payouts among group of cooperative players based on their contributions to the total payout in game theory. In deep learning, SHapley Additive exPlanations (SHAP) value \cite{SHAP} considers each feature as a player and the model prediction as the total payout to estimate feature contributions. Let $\mathcal{M} = \{z_{m_1},...,z_{m_M}\}$ represent the set of all modality features, $\mathcal{S}$ denote the subset, and $V$ symbolize the model outcome. Here, we use features instead of raw data inputs since features are utilized in active learning. To estimate the Shapley value of $i^{th}$ modality feature $z_{m_i}$, we compute the marginal contribution to the subset $\mathcal{S}$ and average over all possible subset selections:

\begin{equation}
    \phi(z_{m_i}) = \sum_{\mathcal{S} \subseteq \mathcal{M} \backslash \{z_{m_i}\}} \frac{|\mathcal{S}|!(|\mathcal{M}| - |\mathcal{S}| - 1)!}{|\mathcal{M}|!} [V(\mathcal{S} \cup \{z_{m_i}\}) - V(\mathcal{S})].
\end{equation}

We use the largest predicted class probability $p_{\hat{y}}$ provided by $f_{mm}$ as the model outcome $V$, where $\hat{y}$ is the pseudo class. For the most common two-modality case, the Shapley values of modality features can be computed as follows ($\emptyset$ represents a zero vector):

\begin{equation}
\begin{aligned}
    &\phi(z_{m_1}) = \frac{1}{2}[V(z_{m_1}, z_{m_2}) - V(\emptyset, z_{m_2}) + V(z_{m_1}, \emptyset) - V(\emptyset, \emptyset)], \\
    &\phi(z_{m_2}) = \frac{1}{2}[V(z_{m_1}, z_{m_2}) - V(z_{m_1}, \emptyset) + V(\emptyset, z_{m_2}) - V(\emptyset, \emptyset)].
\end{aligned}
\end{equation}

The Shapley value could be positive, negative or zero. While the sign indicates in which direction of each modality contributes, our primary interest lies in the extend of its contribution. Hence, we define modality contribution as follows:

\begin{equation}
    \Phi_{m_i} = \frac{\left|\phi(z_{m_i})\right|}{\sum_{i=1}^{M}\left|\phi(z_{m_i})\right|}.
\end{equation}

\subsection{Proposed Method}

Following the proposed guidelines, we redesign the BADGE for multimodal classification scenarios with two modalities, $m_1$ and $m_2$, to achieve more balanced data selection. The $i^{th}$ row of gradient embedding in Eq. \ref{eq:3} could be derived as concatenation of two unimodal gradient embeddings:

\begin{equation}
    g_i = (f_i - 1_{\hat{y}_{mm} = i})z_{m_1} \oplus (f_i - 1_{\hat{y}_{mm} = i})z_{m_2}.
\end{equation}

We then design two weights $w_{m_1}$ and $w_{m_2}$, and scale each unimodal gradient embedding by them respectively:

\begin{equation}
\centering
\begin{aligned}
    w_{m_1}  = \begin{cases}
      1 & \text{if } \Phi_{m_1} \geq \Phi_{m_2} \\
      1-\rho  & \text{if } \Phi_{m_2} > \Phi_{m_1}
        \end{cases} \\
     w_{m_2}  = \begin{cases}
      1-\rho & \text{if } \Phi_{m_1} \geq \Phi_{m_2} \\
      1  & \text{if } \Phi_{m_2} > \Phi_{m_1}
        \end{cases}
\end{aligned}
\end{equation}

\begin{equation}
    g'_i  = w_{m_1}(f_i - 1_{\hat{y}_{mm} = i})z_{m_1} \oplus w_{m_2}(f_i - 1_{\hat{y}_{mm} = i})z_{m_2}.
\end{equation}

Here, $\rho=|\Phi_{m_1} - \Phi_{m_2}|$ is the difference between contributions of two modalities. Note that the gradient embedding of larger l2 norm will be selected more easily by K-Means++ algorithm \cite{badge-iclr/AshZK0A20}. Therefore, by multiplying with these weights, the magnitude of gradient embedding will be suppressed more if their unimodal contributions are more unbalanced. It aligns with our Guideline 1 where we want to punish the samples with unbalanced contributions.

Moreover, we observe that the average $\rho$ of the subset in which the weaker modality dominates is smaller than that of the subset where the stronger modality dominates. See \textbf{Figure \ref{fig:scale}} and our discussion in \textbf{Sec \ref{sec:4.5}}. If $m_1$ is the weaker modality regarding the entire dataset, then we will have $\frac{1}{|X_1|}\sum_{x\in X_1}\rho(x) < \frac{1}{|X_2|}\sum_{x\in X_2}\rho(x)$ for two subsets $X_1$ and $X_2$ dominated by $m_1$ and $m_2$ respectively. It means that the subset where the stronger modality dominates will be suppressed more, and it follows our Guideline 2 to punish the stronger modality on the dataset-level.

Finally, the Guideline 3 is also adhered to. For each sample, the modality with a higher contribution to the model outcome is always assigned a greater weight, resulting in a higher magnitude of unimodal gradient embedding. This ensures that the contribution to data selection is proportional to the contribution to the model outcome and model optimization if selected.

In the end, we perform K-Means++ over the scaled gradient embedding to select candidates for labeling. As a result, our BMMAL strategy could achieve more balanced active learning on multimodal classification than BADGE. It could prevent biased selection towards either the stronger or weaker modalities, thus benefiting multimodal learning.

\section{Experiment}
\label{sec.4}

\subsection{Dataset}

\textbf{Food101} \cite{DBLP:conf/icmcs/WangKTCP15} is a multi-class food recipe dataset with 101 kinds of food. Each recipe consists of a food image and textual recipe description. The dataset consists of 45,719 samples for training and 15,294 samples for testing.

\textbf{KineticsSound} \cite{DBLP:conf/iccv/ArandjelovicZ17} is a sub-dataset containing 31 action classes selected from Kinetics-400 \cite{DBLP:journals/corr/KayCSZHVVGBNSZ17}. These action classes are considered to be correlated to both visual and auditory content. This dataset contains 14,739 clips for training and 2,594 clips for testing.

\textbf{VGGSound} \cite{DBLP:conf/icassp/ChenXVZ20} is a large-scale video dataset with 309 classes. Each video clip is 10-second and captures the object making the sound. We are only able to download 180,911 clips for training and 14,843 clips for testing due to the unavailability of YouTube videos.

\begin{figure}
  \centering
  \includegraphics[width=1.0\linewidth]{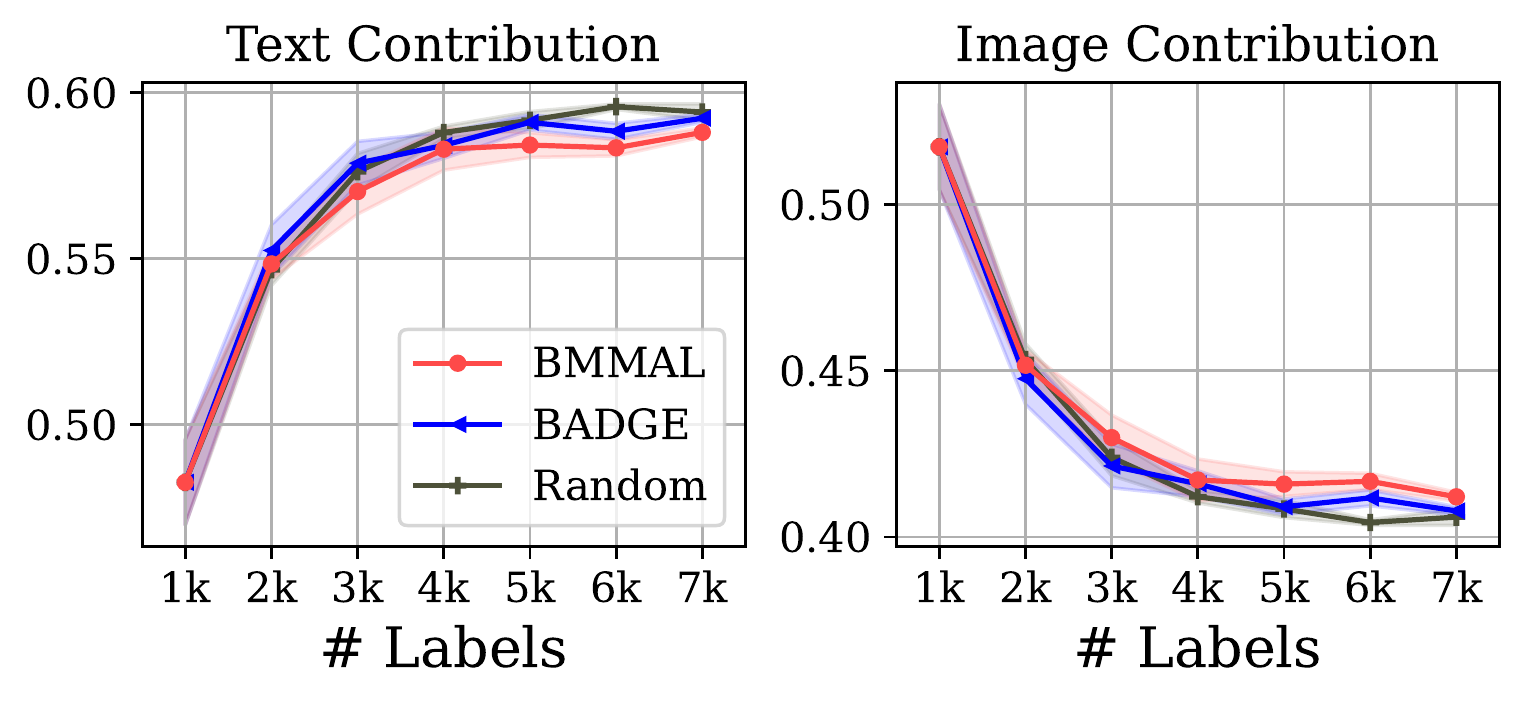}
  \caption{Modality contribution $\Phi$ across different AL iterations on the Food101 test set.}
  \label{fig:food101-contribution}
\end{figure}
\begin{figure}
  \centering
  \includegraphics[width=1.0\linewidth]{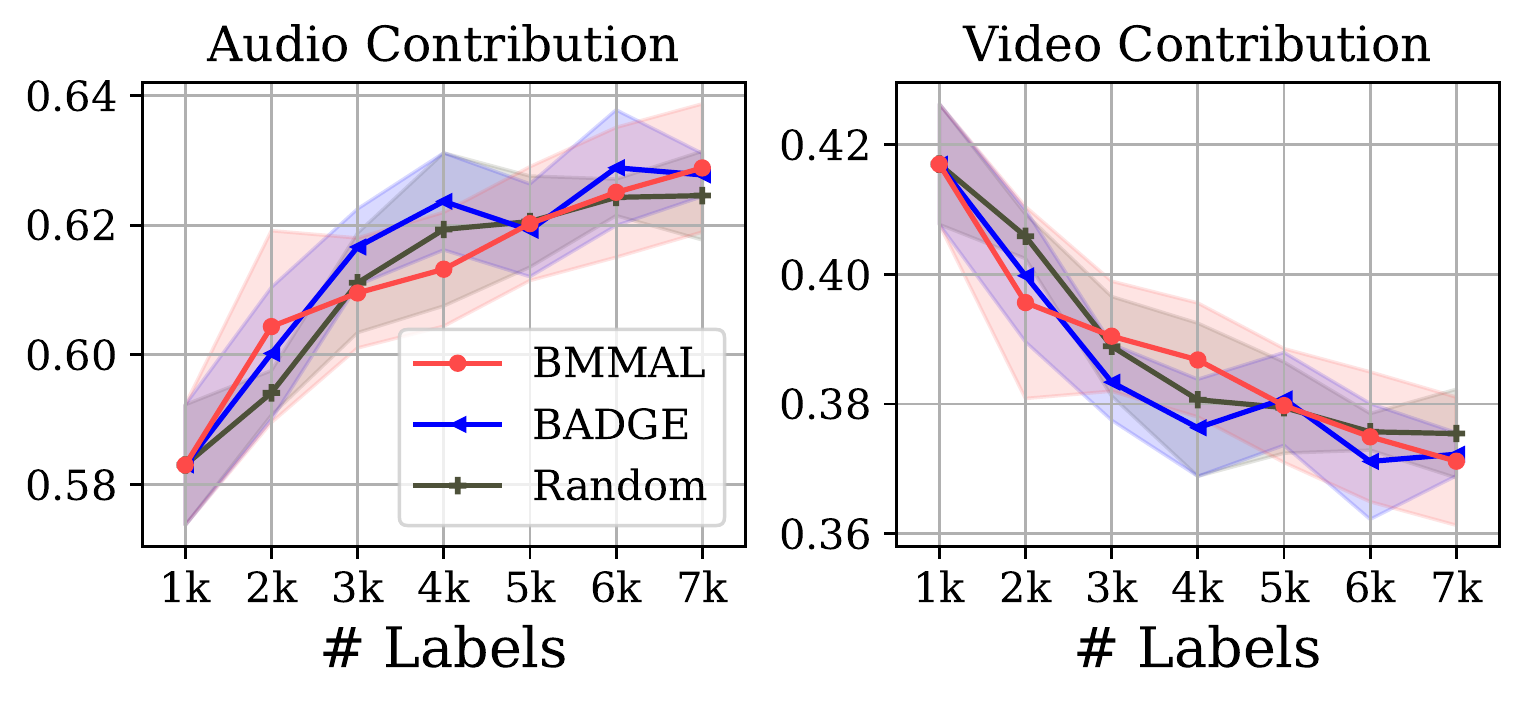}
  \caption{Modality contribution $\Phi$ across different AL iterations on the KineticsSound test set.}
  \label{fig:kinetics_sound_2m-contribution}
\end{figure}

\subsection{Baseline} 
We consider seven existing active learning strategies as baselines. \textbf{Random} selects the data samples randomly from the unlabeled data pool. \textbf{Entropy} \cite{DBLP:series/synthesis/2012Settles} selects data samples with the highest entropy of multimodal classification probabilities. \textbf{CoreSet} \cite{Coreset-SenerS18} filters out a subset of unlabeled data with representative multimodal features via K-center greedy algorithm. \textbf{BADGE} \cite{badge-iclr/AshZK0A20} is a hybrid method that selects diverse data samples by K-means++ sampler over their gradient embedding of multimodal classifier. \textbf{BALD} \cite{BALD} is a Bayesian method to evaluate the mutual information between model predictions and model parameters. Since our model is static, we run five rounds of model forwarding with enabled dropout to obtain the entropy of model parameters. \textbf{DeepFool} \cite{Adversarial-deepfool} adopts an adversarial-like approach that adds small perturbations over multimodal features and selects data whose predictions are flipped. \textbf{GCNAL} \cite{GCNAL} learns an extra graph convolution network to distinguish labelled and unlabelled samples and selects unlabelled samples that are sufficiently different from labelled ones.

\subsection{Experiment Setting}

\textbf{Image-text Classification:} For the Food101 dataset, we adopt ResNet-101 pre-trained on ImageNet as the image backbone and pre-trained Bert-base model \cite{DBLP:conf/naacl/DevlinCLT19} as the text backbone. All unimodal and multimodal classifiers are single FC layers. We use AdamW \cite{adamw} as the optimizer and train the model for 15 epochs in each AL round and adopt random crop, random horizontal flip and random grey scale for image augmentation. 

\textbf{Video Classification:} For VGGSound and KineticsSound, we utilize ResNet2P1D-18 \cite{DBLP:conf/cvpr/TranWTRLP18} as visual backbone. The difference is that it is pre-trained on Kinetics-400 for VGGSound, while it is randomly initialized for KineticsSound. We use the randomly initialized ResNet-18 as an auditory backbone whose input channel is modified from 3 to 1. The video is uniformly sampled into 10 frames at the rate of one frame per second. The audio clip is transformed into a spectrogram with a window length of 512 and an overlap length of 353. For video augmentation, we randomly sample 5 frames out of 10 frames and apply image augmentation techniques on each frame. For audio augmentation, we randomly extract a 5-second audio fragment from the whole audio clip. We use Adam as optimizer and train the model for 45 epochs in each round.

\begin{figure*}
\centering
\begin{subfigure}{.90\textwidth}
  \centering
   \includegraphics[width=0.9\linewidth]{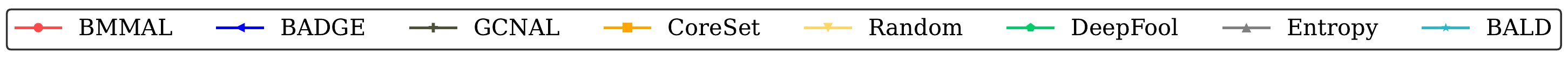}
\end{subfigure}
\begin{subfigure}{.33\textwidth}
  \centering\captionsetup{width=.9\linewidth}
   \includegraphics[width=1.0\linewidth]{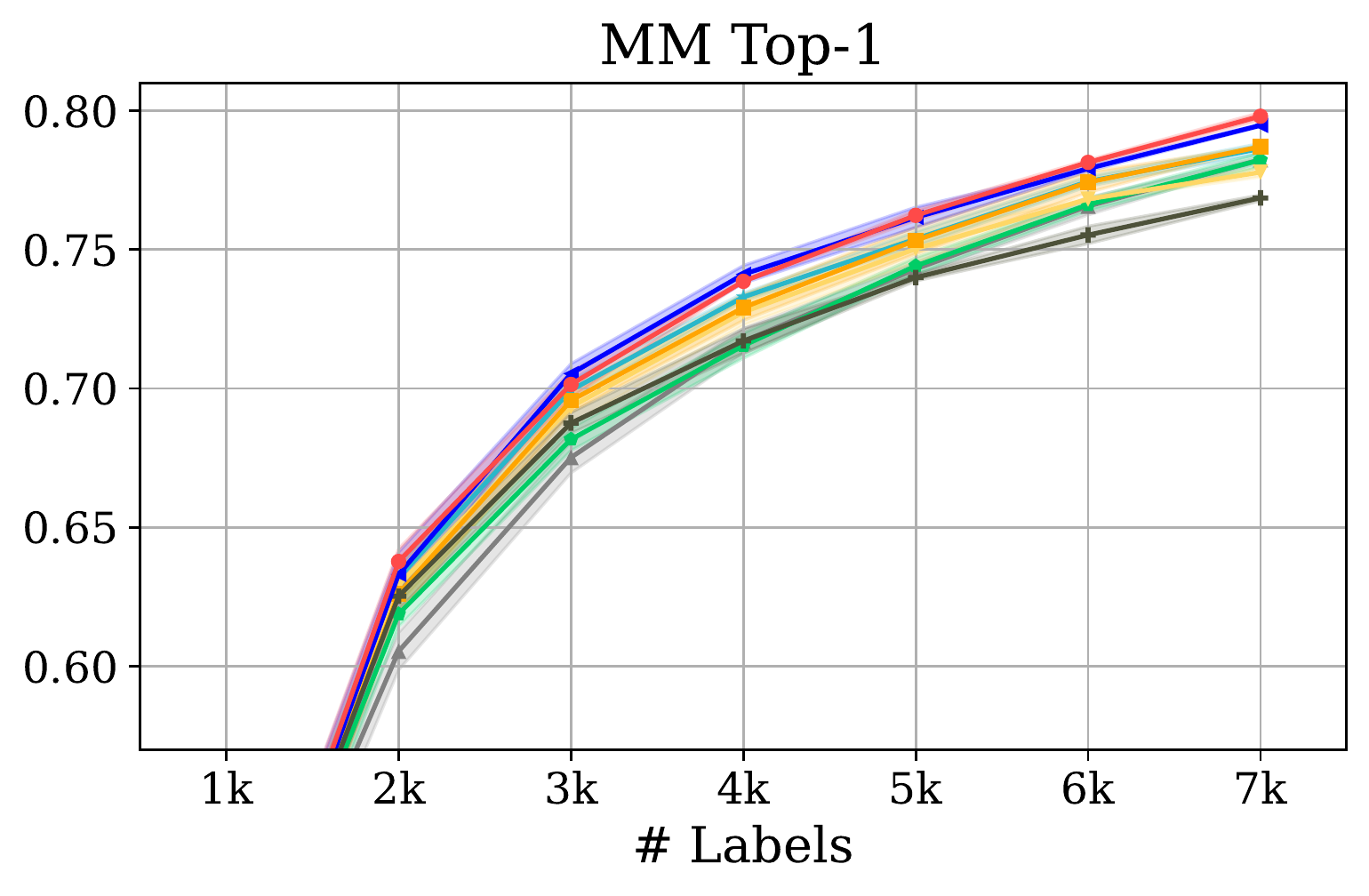}
  \caption{Multimodal performance comparison across AL iterations on Food101.}
   \label{fig:food101-percentage-Top-1-mm}
\end{subfigure}
\begin{subfigure}{.33\textwidth}
  \centering\captionsetup{width=.9\linewidth}
   \includegraphics[width=1.0\linewidth]{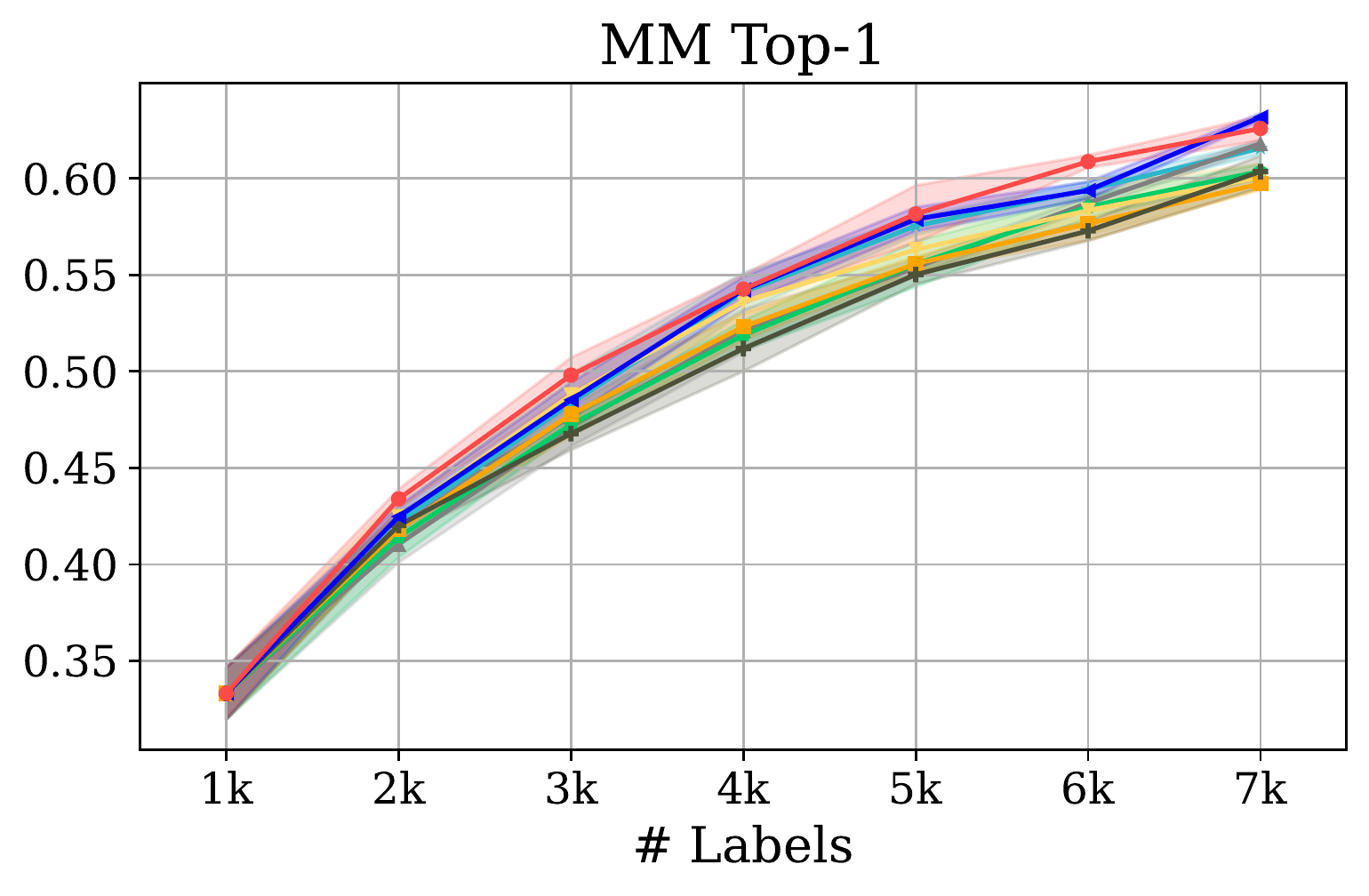}
  \caption{Multimodal performance comparison across AL iterations on KineticsSound.}
   \label{fig:ks-percentage-Top-1-mm}
\end{subfigure}
\begin{subfigure}{.33\textwidth}
  \centering\captionsetup{width=.9\linewidth}
   \includegraphics[width=1.0\linewidth]{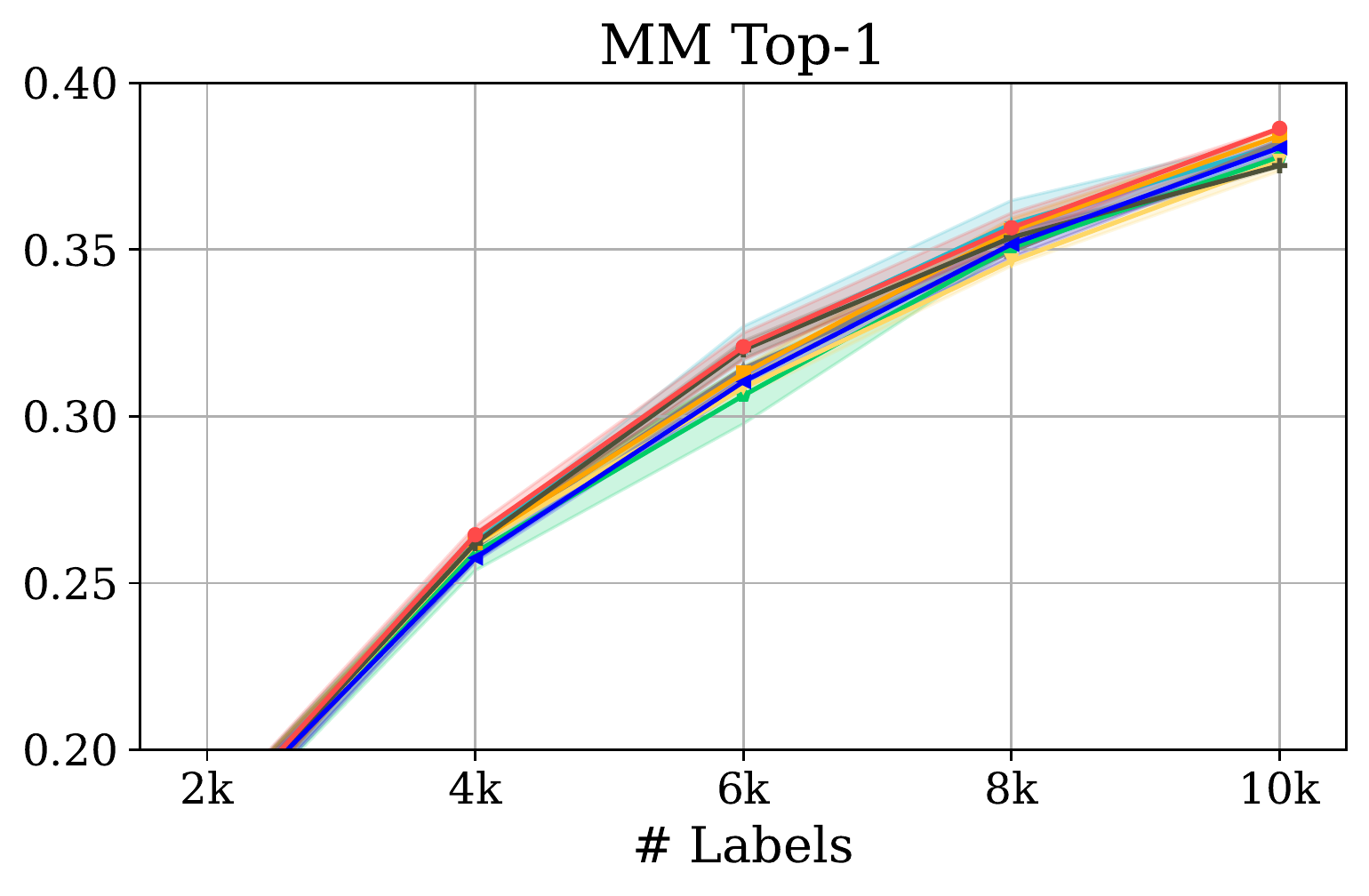}
  \caption{Multimodal performance comparison across AL iterations on VGGSound.}
   \label{fig:vggsound-percentage-Top-1-mm}
\end{subfigure}
\begin{subfigure}{.33\textwidth}
  \centering\captionsetup{width=.9\linewidth}
   \includegraphics[width=1.0\linewidth]{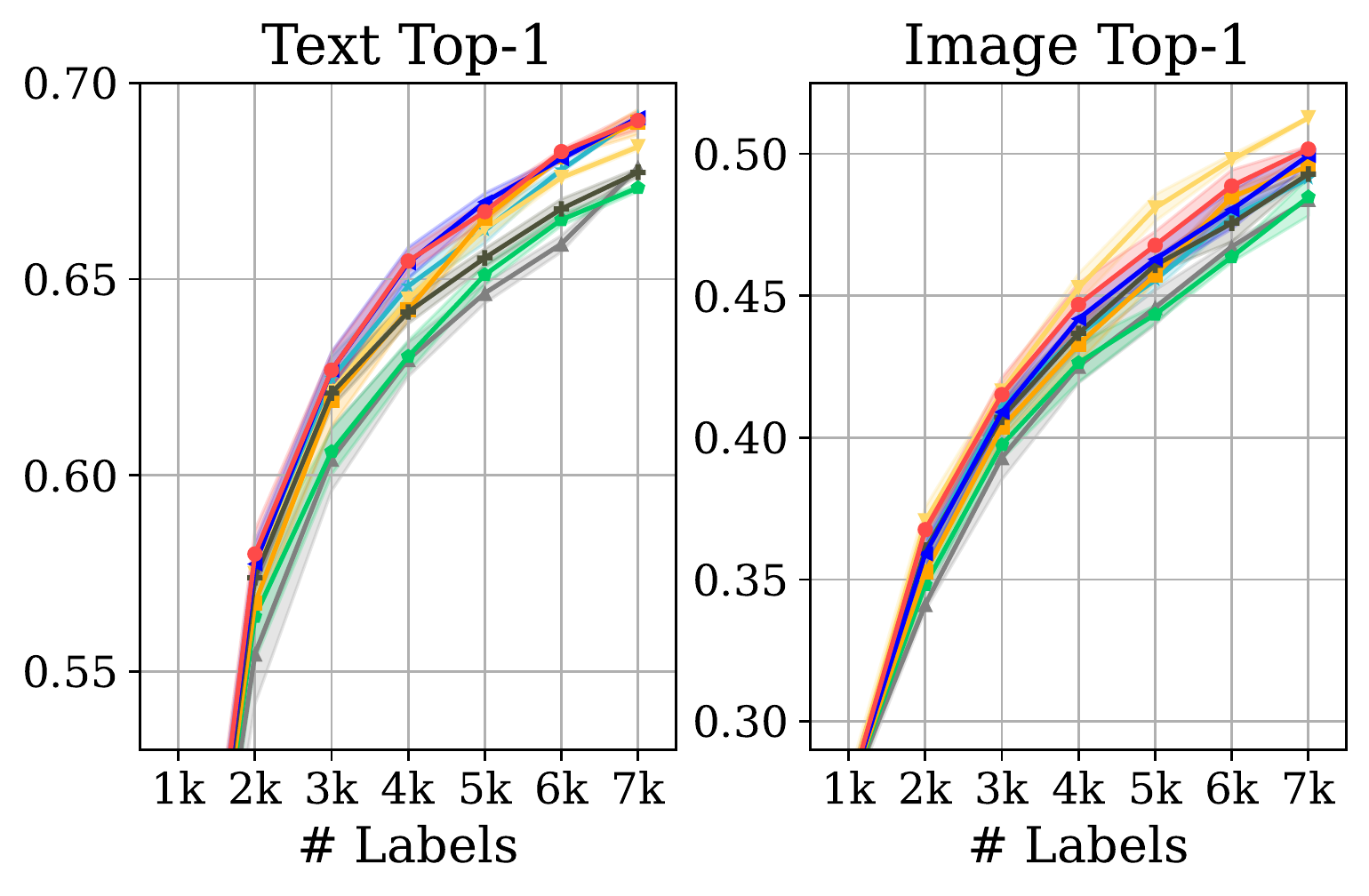}
   \caption{Unimodal performance comparison across AL iterations on Food101.}
   \label{fig:food101-percentage-Top-1-uni}
\end{subfigure}
\begin{subfigure}{.33\textwidth}
  \centering\captionsetup{width=.9\linewidth}
   \includegraphics[width=1.0\linewidth]{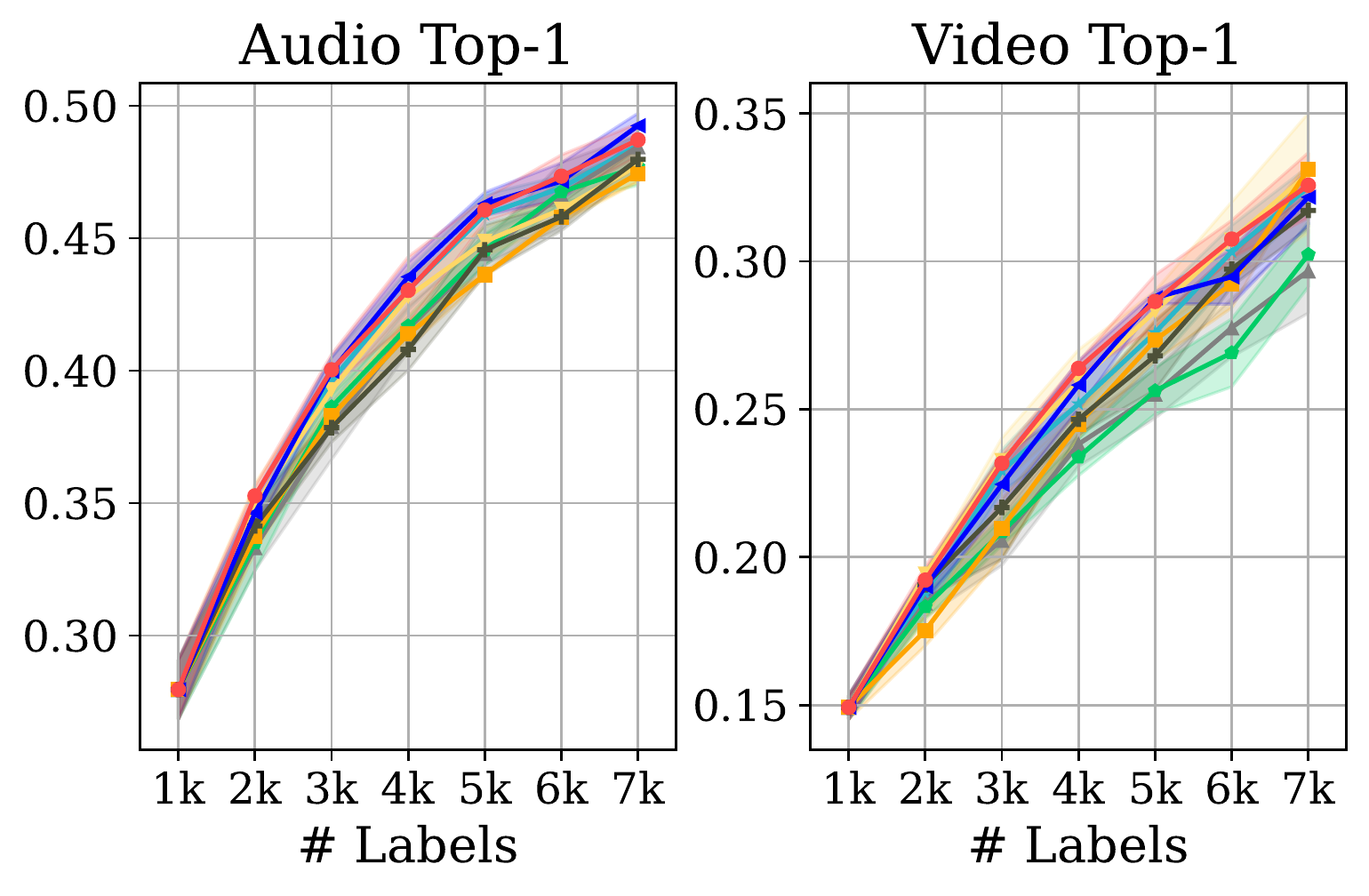}
   \caption{Unimodal performance comparison across AL iterations on KineticsSound.}
   \label{fig:ks-percentage-Top-1-uni}
\end{subfigure}
\begin{subfigure}{.33\textwidth}
  \centering\captionsetup{width=.9\linewidth}
   \includegraphics[width=1.0\linewidth]{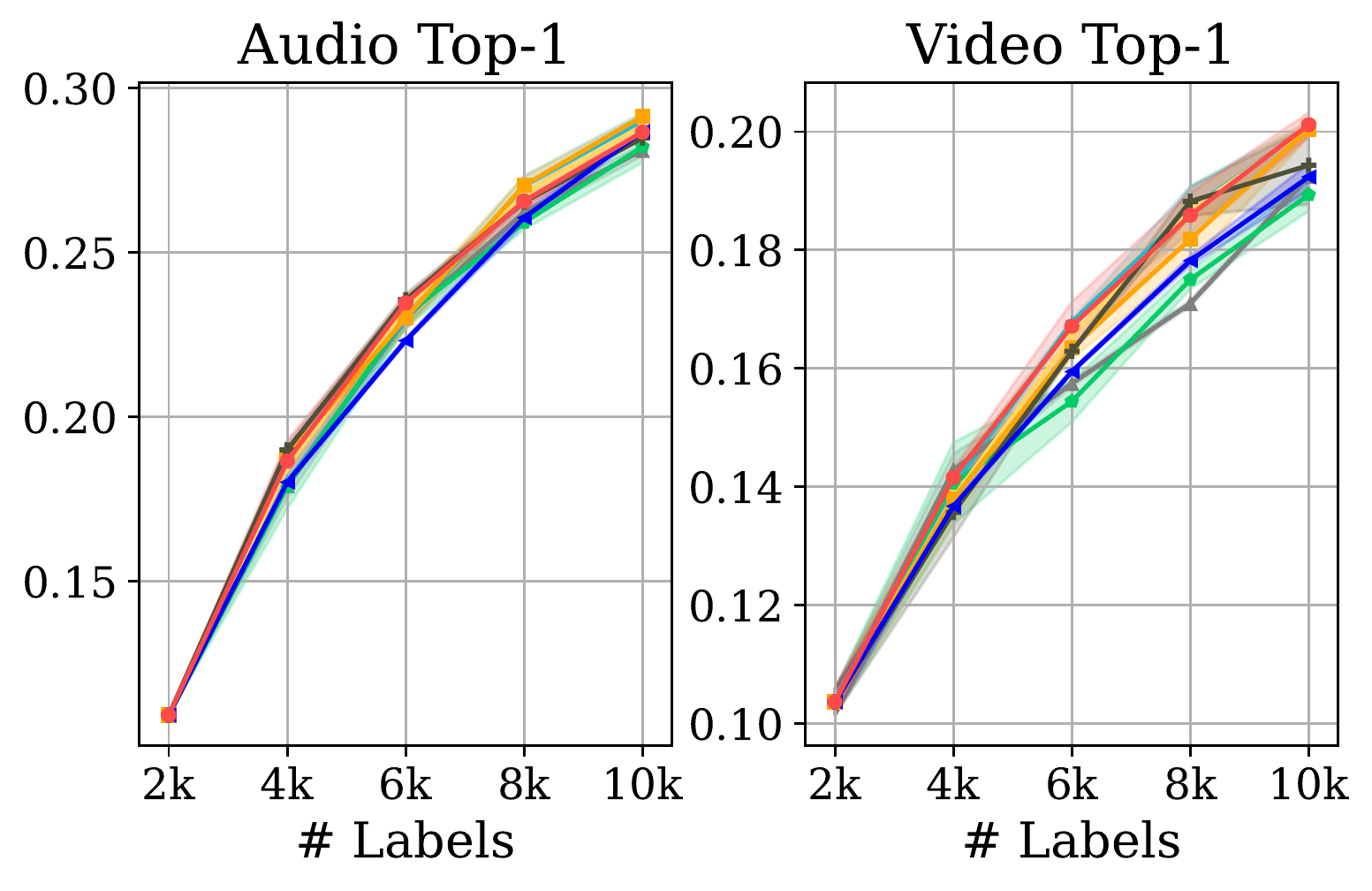}
   \caption{Unimodal performance comparison across AL iterations on VGGSound.}
   \label{fig:vggsound-percentage-Top-1-uni}
\end{subfigure}
\caption{Performance comparison between proposed method and other conventional AL strategies with concatenation fusion method. The metric selected is top-1 accuracy (Top-1) on mulitmodal and unimodal classification.}
\label{fig:image-text-classification-percentage-Top-1}
\end{figure*}

The experiment is repeated 5 times for image-text classification and 3 times for video classification to remove the randomness of the initial querying. For multimodal fusion, we apply concatenation which is a widely used multimodal fusion mechanism on all tasks. In addition, we implement summation and NL-gate \cite{DBLP:conf/cvpr/0004GGH18} that is similar to multi-head attention \cite{DBLP:conf/nips/VaswaniSPUJGKP17} in further experiments.

\subsection{AL Performance}\label{sec.4.4} A fair and good AL strategy ought to select important multimodal data  that could contribute to multimodal tasks and, simultaneously, pay fair attention to weaker modalities and strong modalities to prevent the trained multimodal network from degenerating into only a good unimodal network. We run conventional active learning strategies along with our proposed method BMMAL on several multimodal datasets, and compare their multimodal and unimodal classification accuracy.

We firstly draw the trend of modality contributions to the predicted probability over the ground truth class on test dataset across different active learning iterations in \textbf{Figure \ref{fig:food101-contribution}} and \textbf{Figure \ref{fig:kinetics_sound_2m-contribution}}. As shown in the figures, the textual modal contributes more than the imagery model on the Food101 after second iteration, and the auditory modal contributes more than the visual modal on the KineticsSound. More importantly, the difference between two unimodal contributions of BMMAL is overall smaller than both BADGE and Random. It means that two modalities contribute more equally in the models trained by the data selected by BMMAL.

The performance comparison of each AL iteration on the Food101 dataset is shown in \textbf{Figure \ref{fig:food101-percentage-Top-1-mm}} and \textbf{\ref{fig:food101-percentage-Top-1-uni}}. Note that textual modality is the stronger modality since iteration 2. Our method outperforms all baselines except BADGE in multimodal classification. In text classification, BMMAL, BADGE and CoreSet achieve good performance. In image classification, our method is superior to most of the baselines except Random. From the above comparison, we can tell that BADGE and CoreSet mainly focus on selecting valuable samples over the stronger text modality and ignore the weaker image modality. Although Random uniformly selects multimodal data without any weighting in image classification, it is considered unfair concerning the text modality. 

The performance comparison of each AL iteration on the KineticsSound dataset is shown in \textbf{Figure \ref{fig:ks-percentage-Top-1-mm}} and \textbf{\ref{fig:ks-percentage-Top-1-uni}}. Note that auditory modality is the stronger modality. Our method outperforms all baselines in multimodal classification. BADGE performs the best on audio classification on many iterations, However, its performance declines on video classification indicating that biased data selection might negatively affect multimodal classification. It shows that BADGE tends to assign more importance to audio modality during data selection and such behavior might negatively affect multimodal joint training. 

The performance comparison of each AL iteration on the VGGSound dataset is shown in \textbf{Figure \ref{fig:vggsound-percentage-Top-1-mm}} and \textbf{\ref{fig:vggsound-percentage-Top-1-uni}}. Note that auditory modality is the stronger modality. Our method outperforms BADGE in not only multimodal classification but also in two unimodal classification by an obvious margin. 

\begin{figure*}
\centering
\begin{subfigure}{.33\textwidth}
  \centering
   \includegraphics[width=1.0\linewidth]{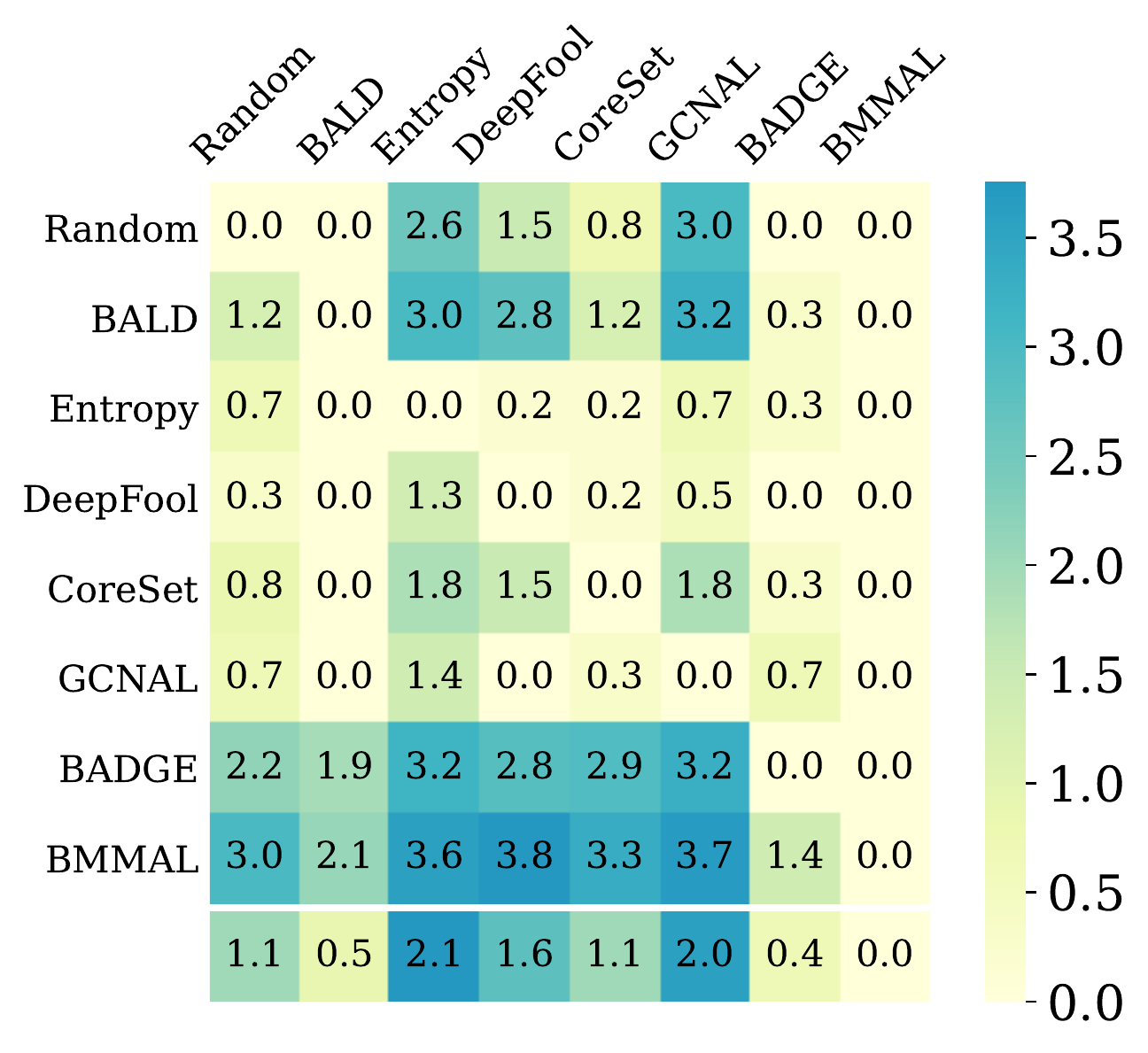}
   \caption{Pairwise comparison on multimodal classification.}
   \label{fig:pairwise-mm}
\end{subfigure}
\begin{subfigure}{.33\textwidth}
  \centering\captionsetup{width=.9\linewidth}
   \includegraphics[width=1.0\linewidth]{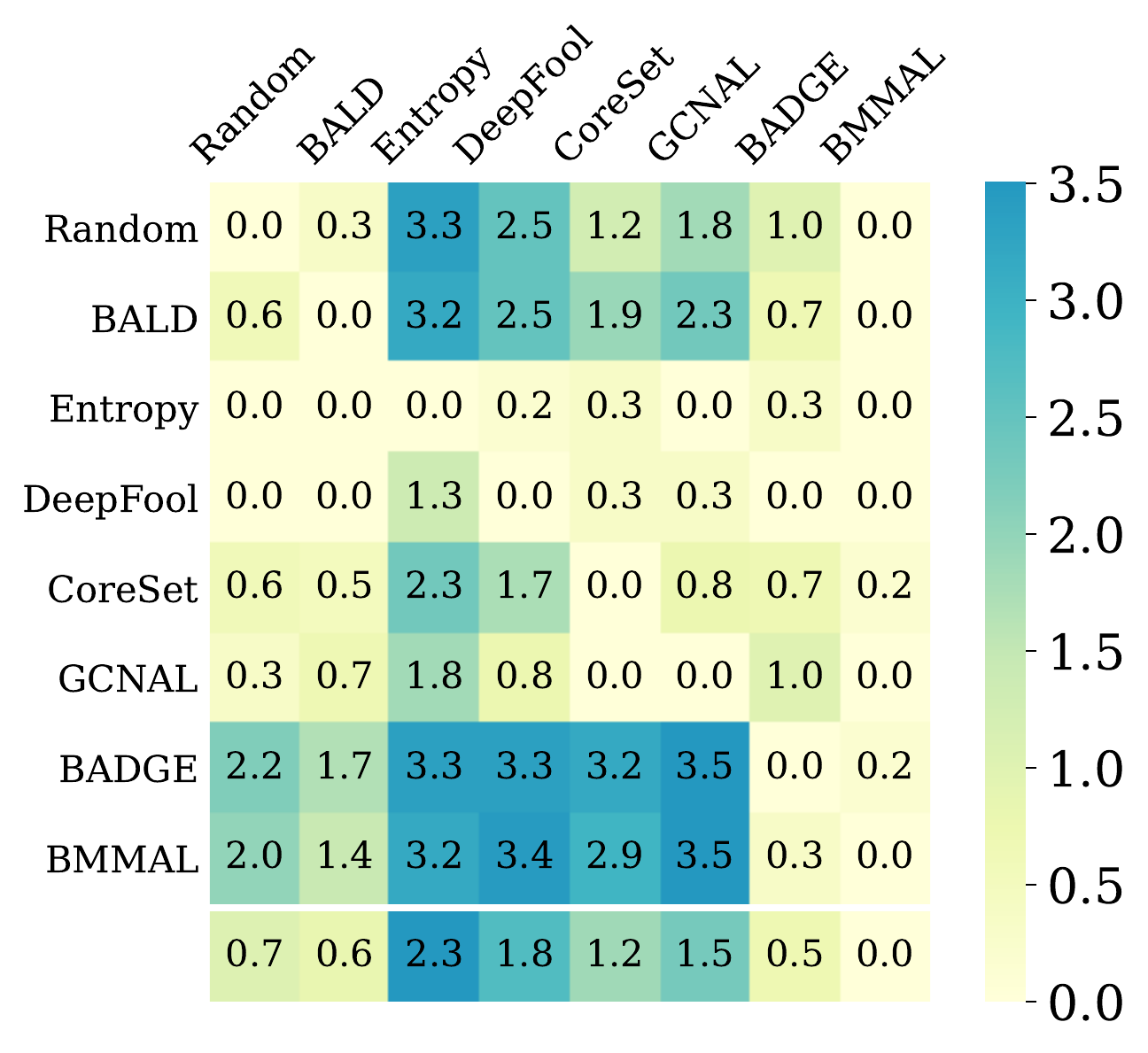}
  \caption{Pairwise comparison on unimodal classification of stronger modalities.}
   \label{fig:pairwise-weak}
\end{subfigure}
\begin{subfigure}{.33\textwidth}
  \centering\captionsetup{width=.9\linewidth}
   \includegraphics[width=1.0\linewidth]{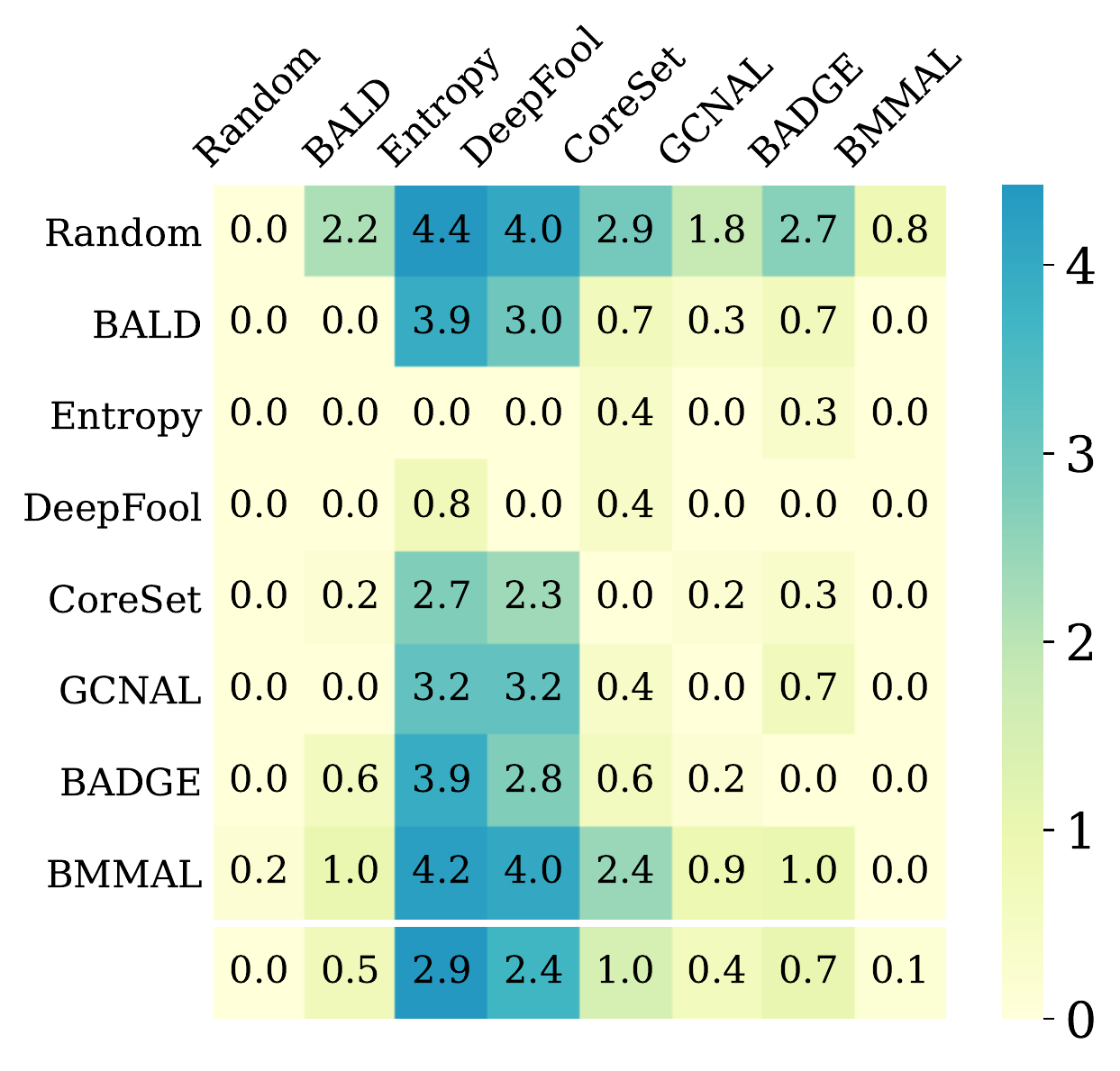}
   \caption{Pairwise comparison on unimodal classification of weaker modalities.}
   \label{fig:pairwise-stron}
\end{subfigure}
\caption{Pairwise comparison of all active learning strategies. Each element in the matrix $P_{i,j}$ represents the number of times strategy $i$ outperforms strategy $j$. A strategy is considered better if its row-wise value is larger, indicating that it beats other strategies more often. On the other hand, a strategy is better if its column-wise value is smaller, meaning it is rarely beaten by other strategies. The maximum value of each cell is 5, which is the total number of experimental settings. The bottom row displays the column-wise average values (lower is better).}
\label{fig:pairwise}
\end{figure*}

\textbf{Findings.} Our first finding is that AL methods such as BADGE and BALD which win at classification of the stronger modality could stand a good chance of failing at classification of the weak modality. This may be due to biased data selection towards the stronger modality, and it is undesirable for balanced multimodal learning. Our second finding is that Random and CoreSet could perform better in the weaker modality, whereas they are inferior in multimodal classification because random selection treats every sample with absolute fairness and CoreSet focuses too much on the weak modality which are both unfair concerning the stronger modality. Finally, our method achieves a fairer multimodal data selection with a better trade-off between weak and strong modalities.

\subsection{Ablation Study}
\label{sec:4.5}

\textbf{Pairwise Comparison.} We illustrate the results across various experimental settings in matrix $P$  in \textbf{Figure \ref{fig:pairwise}} \cite{badge-iclr/AshZK0A20}. We compute the t-score for each repeated experiment and use the two-sided t-test to compare the performance of paired strategies on the test set with a 0.9 confidence interval. If strategy $i$ significantly outperforms strategy $j$, we add $1/L$ to $P_{i,j}$, where $L$ is the total number of iterations for a single experiment setting. The maximum cell value equals the total number of experiment settings. $P_{i,j}$ indicates the number of times strategy $i$ significantly outperforms strategy $j$. We compute the matrix for both multimodal and unimodal classification for stronger (text for Food101, audio for KineticsSound and VGGSound) and weaker modalities (image for Food101, video for KineticsSound and VGGSound). The three matrices demonstrate that our proposed method outperforms most baselines across settings. Specifically, BMMAL surpasses BADGE in multimodal classification and unimodal classification on weaker modalities, while performing comparably with BADGE in unimodal classification on stronger modalities. This suggests that the performance improvement of BMMAL in multimodal classification mainly stems from enhancing weaker modalities while maintaining stable performance in stronger modalities.

\begin{figure}
\centering
\begin{subfigure}{0.4\textwidth}
  \centering\captionsetup{width=.9\linewidth}
   \includegraphics[width=1.0\linewidth]{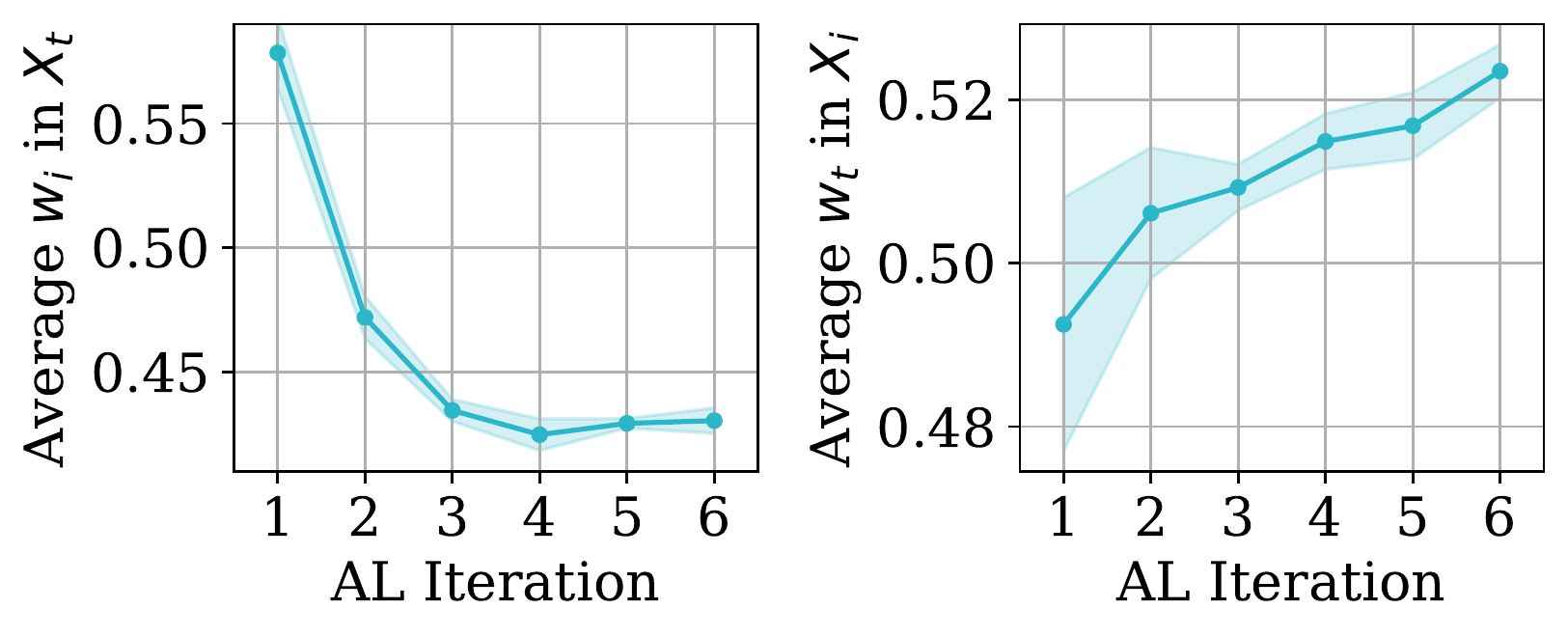}
  \caption{Average weight $w_i$ in $X_t$ and average weight $w_t$ in $X_i$ on the Food101 dataset.}
   \label{fig:food101-scale}
\end{subfigure}
\begin{subfigure}{0.4\textwidth}
  \centering\captionsetup{width=.9\linewidth}
   \includegraphics[width=1.0\linewidth]{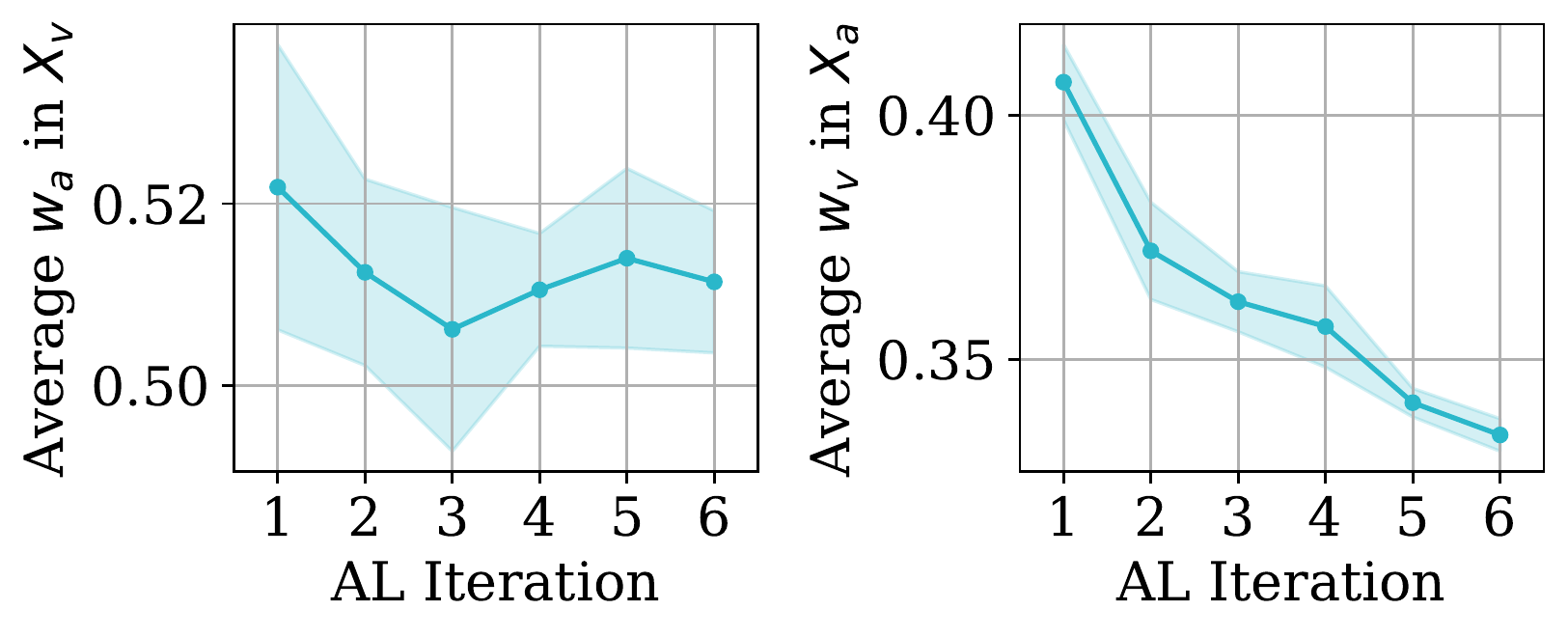}
   \caption{Average weight $w_a$ in $X_v$ and average weight $w_v$ in $X_a$ on the KineticsSound dataset.}
   \label{fig:ks-scale}
\end{subfigure}
\caption{Average weight for the weaker modality in a sub-dataset dominated by the other stronger modality.}
\label{fig:scale}
\end{figure}

\textbf{Dominance Degree}. As described in \textbf{Eq. \ref{eq:6}}, we divide the entire unlabeled dataset into multiple sub-datasets in which modality $m_i$ contributes the most. The Food101 dataset is divided into $X_t$ and $X_i$ dominated by text and image modality, respectively. In \textbf{Figure \ref{fig:food101-scale}}, the average weight values of the weaker modality are showed. As shown before in \textbf{Figure \ref{fig:food101-contribution}}, text modality is the stronger one starting from the second iteration.  The average value of $w_i$ in $X_t$ accordingly becomes less than that of $w_t$ in $X_i$ from the second iteration, meaning that the average difference value $\rho$ between two unimodal contributions in $X_t$ is larger than in $X_i$. The KineticsSound dataset is divided into $X_v$ and $X_a$ dominated by video and audio modality, respectively. In \textbf{Figure \ref{fig:ks-scale}}, the average weight values of the weaker modality are showed. Similarly, the average difference value $\rho$ between two unimodal contributions in $X_a$ is larger than in $X_v$. Consequently, on the dataset-level, the sub-dataset dominated by the weaker modality receives less punishment compared to the sub-dataset dominated by the stronger modality.

\begin{figure}[t]
  \centering
   \includegraphics[width=0.90\linewidth]{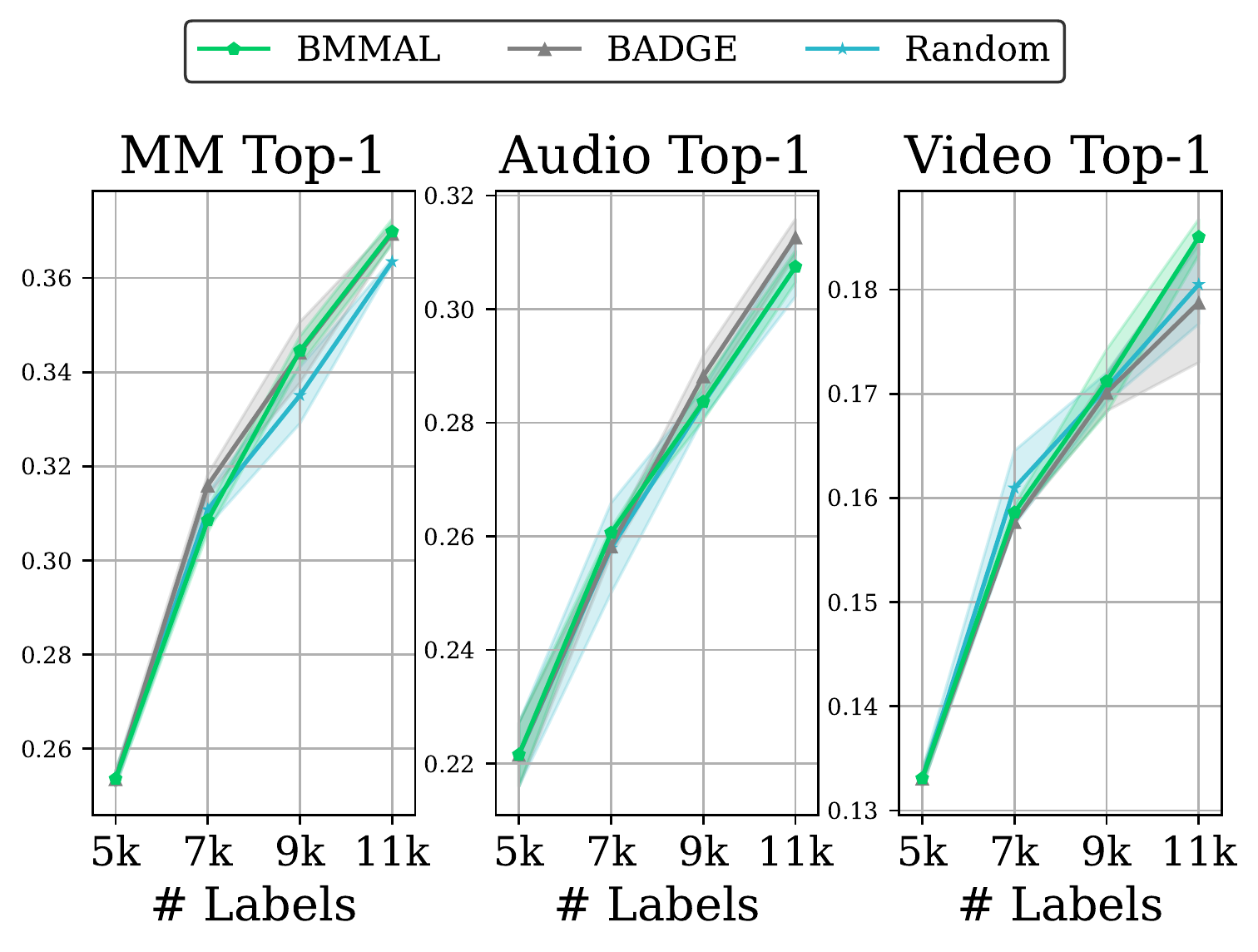}
   \caption{Multimodal and unimodal classification performance comparison with NL-gate fusion method on the VGGSound dataset.}
   \label{fig:vggsound-nlgate}
\end{figure}

\textbf{Different Fusion Mechanisms.} We perform experiment by changing the fusion method from concatenation into summation on Food101 and KineticsSound, while keeping other settings unchanged. We include the performance comparison in the pairwise comparison and present the iterative comparison in the supplementary materials. Furthermore, we change concatenation to NL-gate for mixing video and audio features on the VGGSound dataset, setting the initial budget to 5,000 and the AL budget for each round to 2,000, as NL-gate requires more data to demonstrate its efficiency in fusion. We provide the implementation details in the supplementary materials. As shown in \textbf{Figure \ref{fig:vggsound-nlgate}}, our method achieves comparable multimodal classification performance to BADGE and becomes worse on auditory classification. However, for the weaker visual classification, our method outperforms the others, demonstrating its effectiveness in balancing weak and strong modalities.

\textbf{Large-scale Active Learning.} We conduct experiment on VGGSound with larger budget size of 5,000 to validate our method on large-scale active learning for multimodal video classification. The results are averaged and shown in \textbf{Table \ref{table:vggsound_full}}. On video classification, the performance of BADGE degrades and becomes worse than random selection, while our method achieves improvement over BADGE and random selection. On audio classification, BADGE and our method are comparable and are both better than random selection. As a result, our method performs better than BADGE and can save around 5k labels compared with random selection if target multimodal classification top-1 accuracy is set to 0.435.

\begin{table}
\centering
\small
\begin{tabular}{|c|c|c|c|c|c|c|}
\hline
                            & \#Labels & 5k    & 10k            & 15k            & 20k            & 25k        \\ \hline
\multirow{3}{*}{MM-Top-1}    & Random  & 0.261 & 0.340          & 0.387          & 0.418          & 0.435      \\ \cline{2-7} 
                            & BADGE    & 0.261 & \textbf{0.355}          & 0.406          & 0.433          & 0.451       \\ \cline{2-7} 
                            & BMMAL    & 0.261 & 0.352          & \textbf{0.407}          & \textbf{0.437}          & \textbf{0.458}      \\ \hline
\multirow{3}{*}{Audio-Top-1} & Random  & 0.189 & 0.251          & 0.295          & 0.318          & 0.334        \\ \cline{2-7} 
                            & BADGE    & 0.189 & \textbf{0.262}          & 0.307          & 0.332          & 0.345           \\ \cline{2-7} 
                            & BMMAL    & 0.189 & 0.261          & \textbf{0.308}          & \textbf{0.333}          & \textbf{0.350}          \\ \hline
\multirow{3}{*}{Video-Top-1} & Random  & 0.145 & 0.178          & 0.203          & 0.220          & 0.229          \\ \cline{2-7} 
                            & BADGE    & 0.145 & \textbf{0.184}          & 0.206          & 0.218          & 0.225          \\ \cline{2-7} 
                            & BMMAL    & 0.145 & 0.180          & \textbf{0.208}          & \textbf{0.222}          & \textbf{0.231}      \\ \hline
\end{tabular}
\caption{AL performance on VGGSound dataset with budget size of 5,000. The best results are highlight in bold.}
\label{table:vggsound_full}
\end{table}

\begin{figure}[t]
  \centering
   \includegraphics[width=1.00\linewidth]{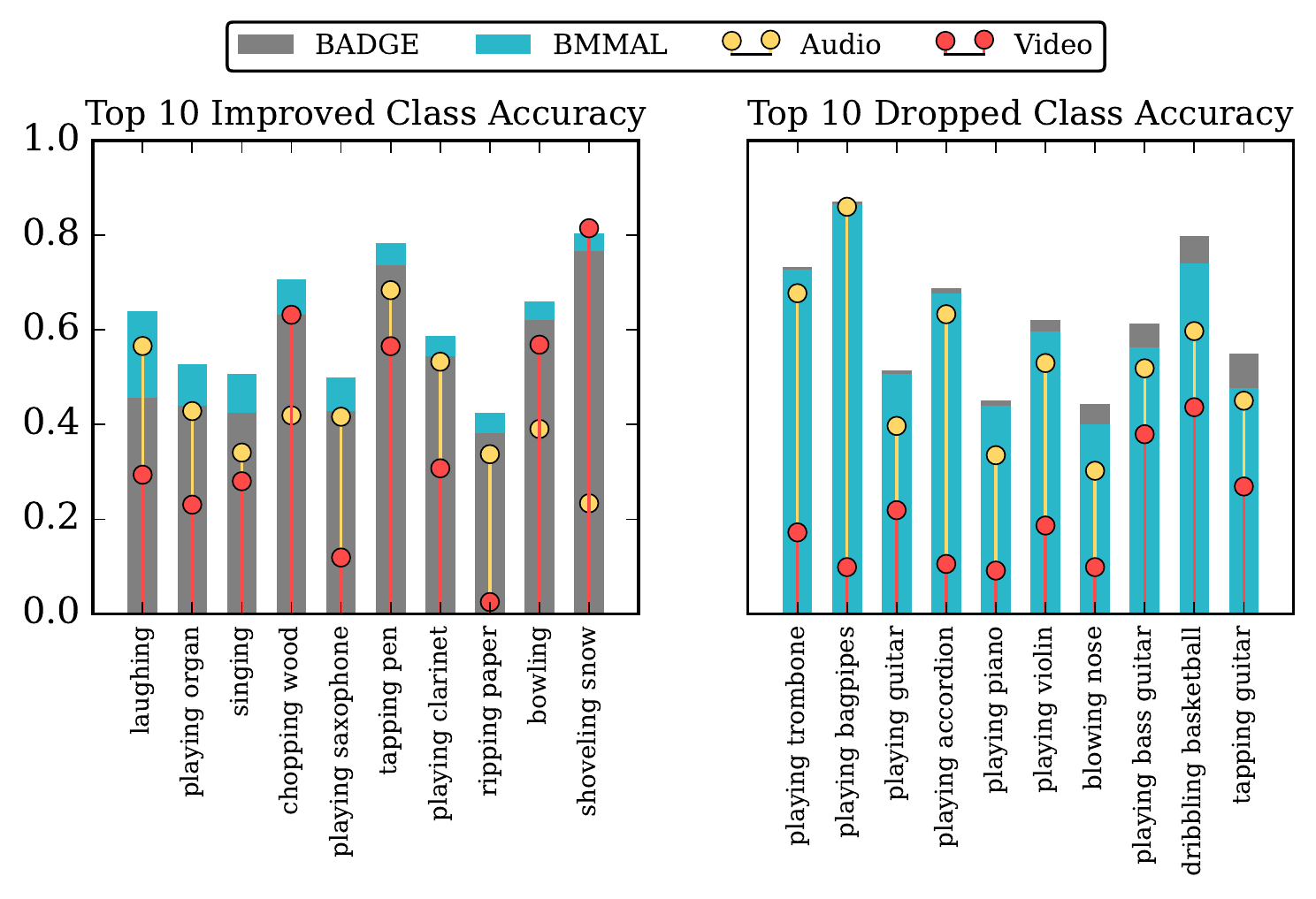}
   \caption{Top 10 improved and dropped classes based on the improvement of BMMAL to BADGE on multimodal classification accuracy on KineticsSound with 5K labeled samples. Bars represent multimodal classification accuracy. Stems represent unimodal classification accuracy.}
   \label{fig:class_improvement}
\end{figure}

\textbf{Classwise Performance Comparison}. We show the classwise performance comparison on the KineticsSound dataset. As shown in \textbf{Figure \ref{fig:class_improvement}}, the gain is more significant than the drop. Moreover, improved classes such as 'chopping wood', 'bowling' and 'shoveling snow' carry more visual information, and dropped classes are mostly dominated by the auditory modality. Note that KineticsSound is a dataset where audio contributes more than vision, which means that BMMAL avoids biased selection over auditory modality and focuses more on the weaker visual modality.

\section{Discussion}
\label{sec.5}

In this paper, we evaluate how existing active learning strategies perform on multimodal classification. Our empirical studies show that they might treat different modalities unfairly, and it could lead to performance degradation for multimodal learning. We propose BMMAL to mitigate this unfairness by separately scaling unimodal gradient embeddings, which avoids mixing all unimodal information and well retain characteristics of each modality. The method performs well on multiple datasets and can be potentially applied on large-scale multimodal active learning. 

% \textbf{Limitation and Future Works}. Although we propose guidelines for designing balanced multimodal active learning, these are lack of theoretical supports and can not guarantee that the redesigned strategy must perform better. Our method is based on gradient embedding, which lacks the flexibility to be applied on more advanced multimodal fusion mechanisms or architectures. The future works ought to investigate the theory for multimodal active learning, develop more advanced methods to obtain better multimodal data selection and inspect multimodal AL from other perspectives such as modality correlation.

\newpage

% \begin{acks}
% \end{acks}

\bibliographystyle{ACM-Reference-Format}
\balance
\bibliography{reference}

\newpage

% \appendix









\appendix

\section{Computational Complexity}

Computing the Shapley values of each unimodal feature requires to perform inference $2^M$ times in total, where $M$ is the number of modalities. In our two-modality learning case, we need to perform inference four times with different combination of unimodal features to obtain the Shapley values, which is acceptable. Then, given the computed gradient embedding of $N$ unlabeled samples, the sampling time complexity of BMMAL is $\mathcal{O}(NBDK)$, where $B$ is the query budget of each AL round, $D$ is the size of weight matrix of the last linear classifier and $K$ is the number of classes.

\section{Implementation of NL-gate}

NL-gate \cite{DBLP:conf/cvpr/0004GGH18} is a mid-fusion mechanism that behaves similar to multi-head attention. We implement it in the video classification task, where Resnet-18 is utilized as the audio backbone and Resnet2P1D-18 is utilized as the video backbone. Note that both Resnet-18 and Resnet2P1D-18 have four blocks. We extract the middle 2D audio features from the third block of Resnet-18 and the middle 3D video features from the third block of Resnet2P1D-18 as inputs to the NL-gate.

\begin{figure}[!b]
  \centering
   \includegraphics[width=0.9\linewidth]{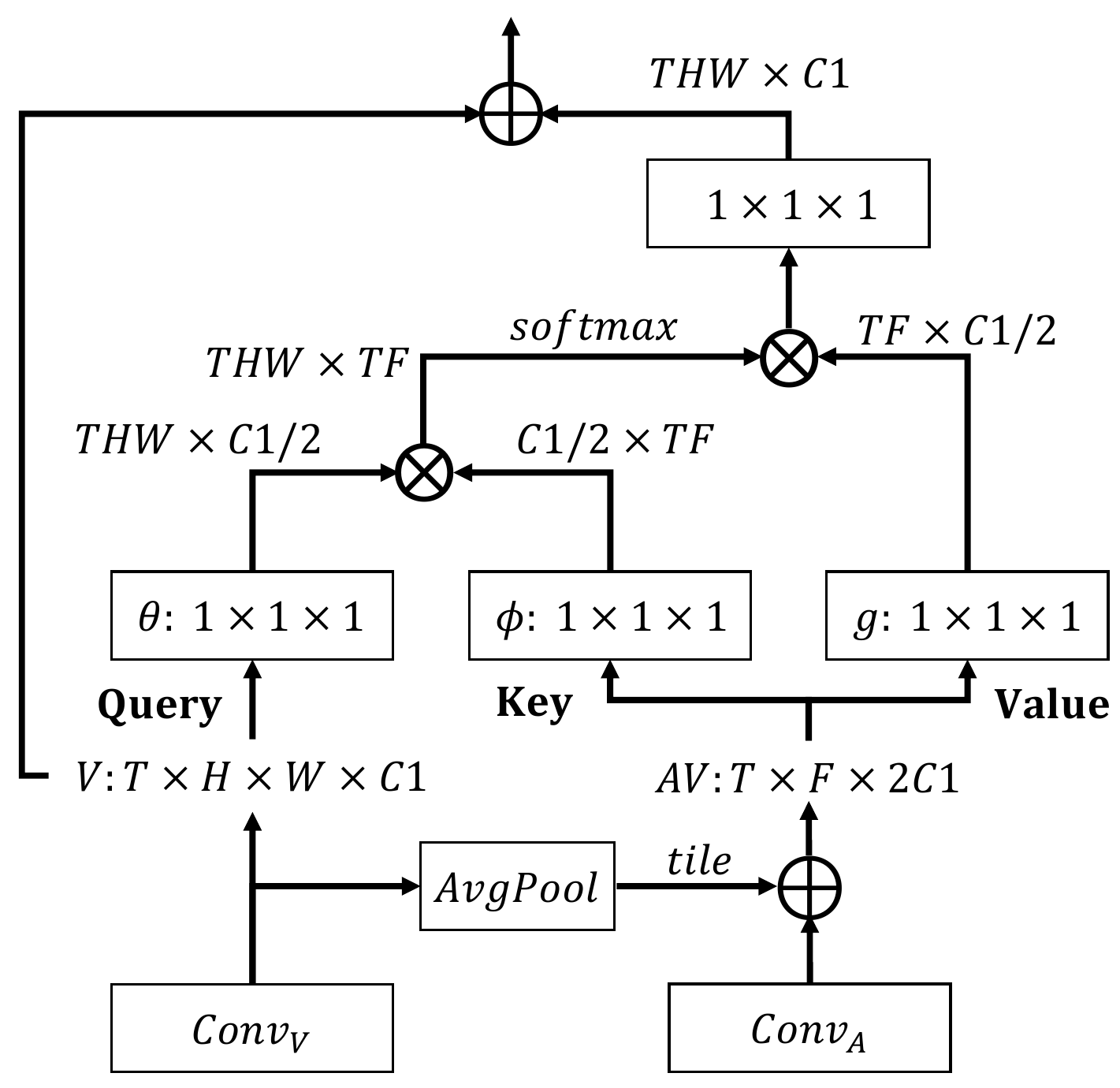}
   \caption{The implementation of NL-gate. We use the 3D video feature as query and the 2D concatenated audio and video feature as key and value.}
   \label{fig:nl-gate}
\end{figure}

We show the implementation of NL-gate in \textbf{Figure \ref{fig:nl-gate}}. The 3D video feature is average pooled over the spatial channels into a 1D video feature. It is then tiled over the frequency channel into a 2D video feature that has the same size as the 2D audio feature. The concatenation of the 2D video feature and the 2D audio feature is used as key and value in NL-gate. The original 3D video feature is used as query in NL-gate. After audio and video features are mixed, they will be processed with a random initialized module with the same layout as the fourth Block of Resnet2P1D-18 to produce the final feature. To compute the marginal unimodal contribution, we choose to compute the Shapley values of the features generated by the last shared convolution layers ($Conv_V$ and $Conv_A$) before the NL-gate fusion module.

\section{Split the Large Unlabeled Data Pool}

In large-scale AL experiments, the gradient embedding produced by all unlabeled data samples could be too large to be stored in the memory. To address this issue, we split the unlabeled data pool into $S$ smaller pools to save memory space, where $S$ is the split size. After splitting, we query $\frac{N}{S}$ unlabeled samples from each smaller pool and aggregate them to form the final query set. The space complexity of BMMAL is correspondingly reduced by $S$ times. Moreover, the sampling time complexity becomes $\mathcal{O}(\frac{N}{S}\frac{B}{S}DKS) = \mathcal{O}(\frac{1}{S}NBDK)$, which is also reduced by $S$ times compared with original time complexity. We use split size of eight in the large-scale AL experiment with the VGGSound-full dataset. Although splitting might affect the AL performance, we observe that both BMMAL and BADGE still perform better than random data selection. It indicates that splitting the unlabeled data pool is acceptable in large-scale AL.

\section{AL Performance with Summation}
\begin{figure*}
\centering
\begin{subfigure}{.90\textwidth}
  \centering
   \includegraphics[width=0.9\linewidth]{figures/legends.pdf}
\end{subfigure}
\begin{subfigure}{.45\textwidth}
  \centering\captionsetup{width=.8\linewidth}
   \includegraphics[width=1.0\linewidth]{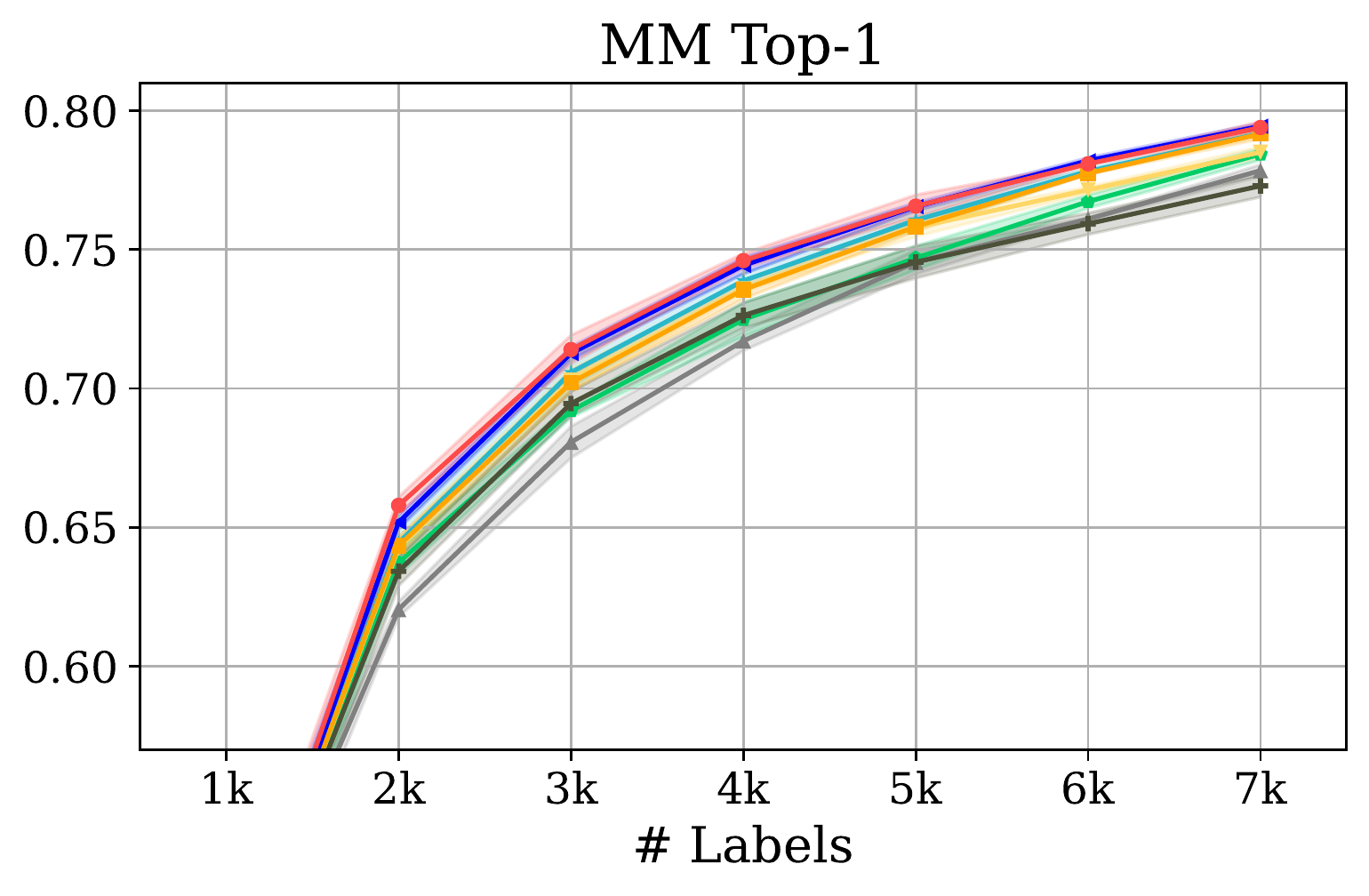}
  \caption{Multimodal performance comparison across AL iterations on Food101.}
   \label{fig:food101-percentage-Top-1-mm}
\end{subfigure}
\begin{subfigure}{.45\textwidth}
  \centering\captionsetup{width=.8\linewidth}
   \includegraphics[width=1.0\linewidth]{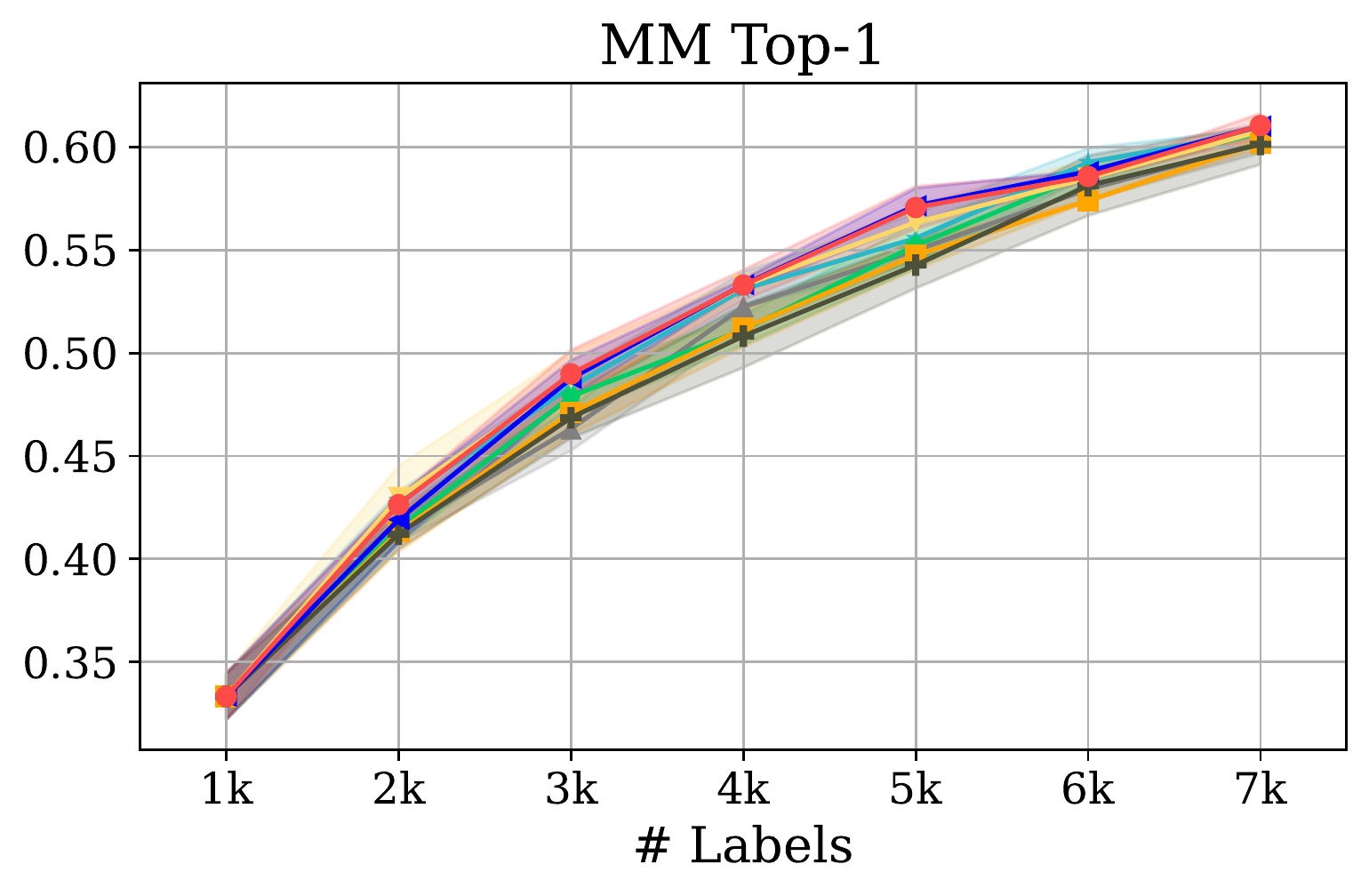}
  \caption{Multimodal performance comparison across AL iterations on KineticsSound.}
   \label{fig:ks-percentage-Top-1-mm}
\end{subfigure}
\begin{subfigure}{.45\textwidth}
  \centering\captionsetup{width=.8\linewidth}
   \includegraphics[width=1.0\linewidth]{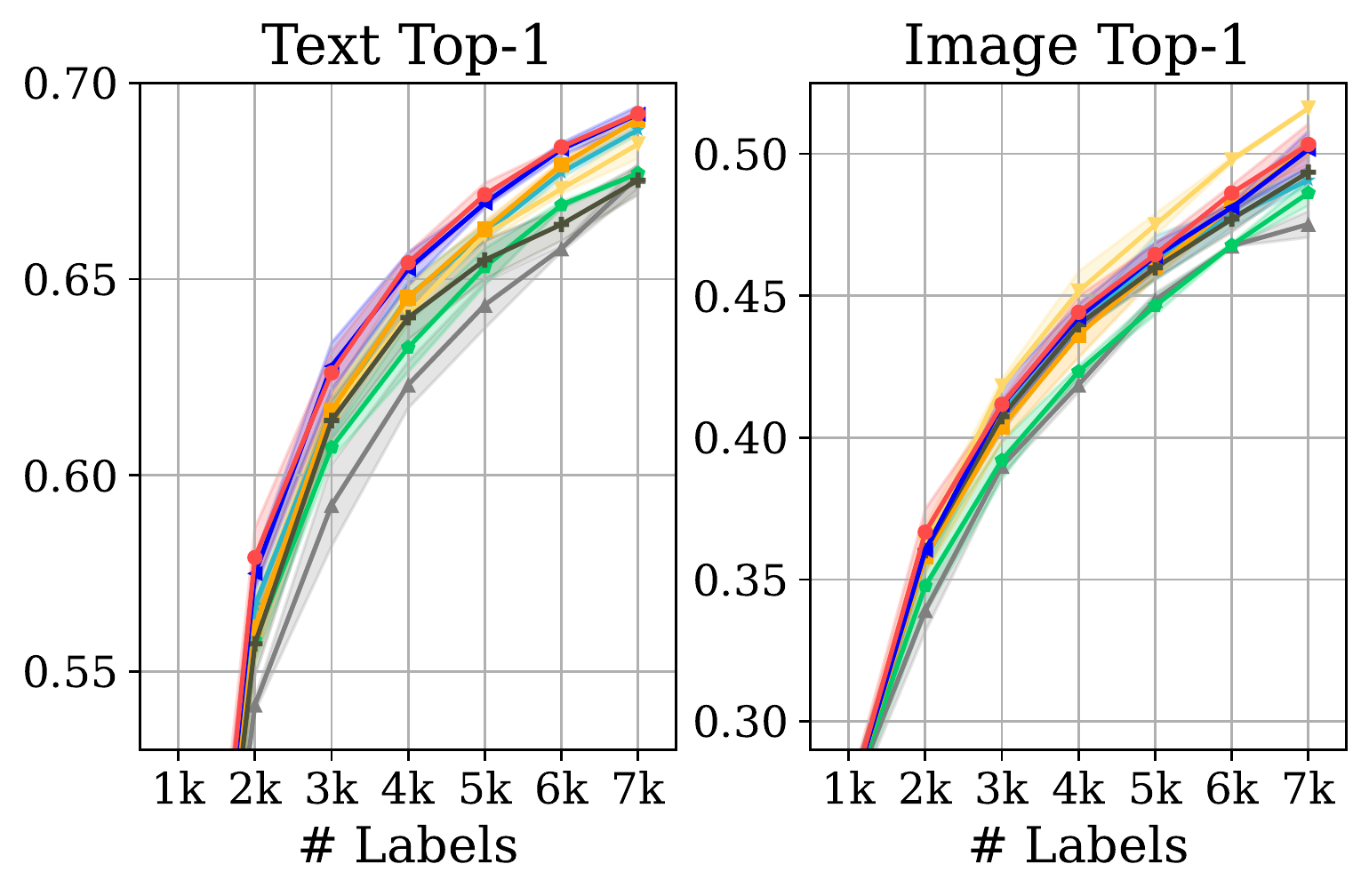}
   \caption{Unimodal performance comparison across AL iterations on Food101.}
   \label{fig:food101-percentage-Top-1-uni}
\end{subfigure}
\begin{subfigure}{.45\textwidth}
  \centering\captionsetup{width=.8\linewidth}
   \includegraphics[width=1.0\linewidth]{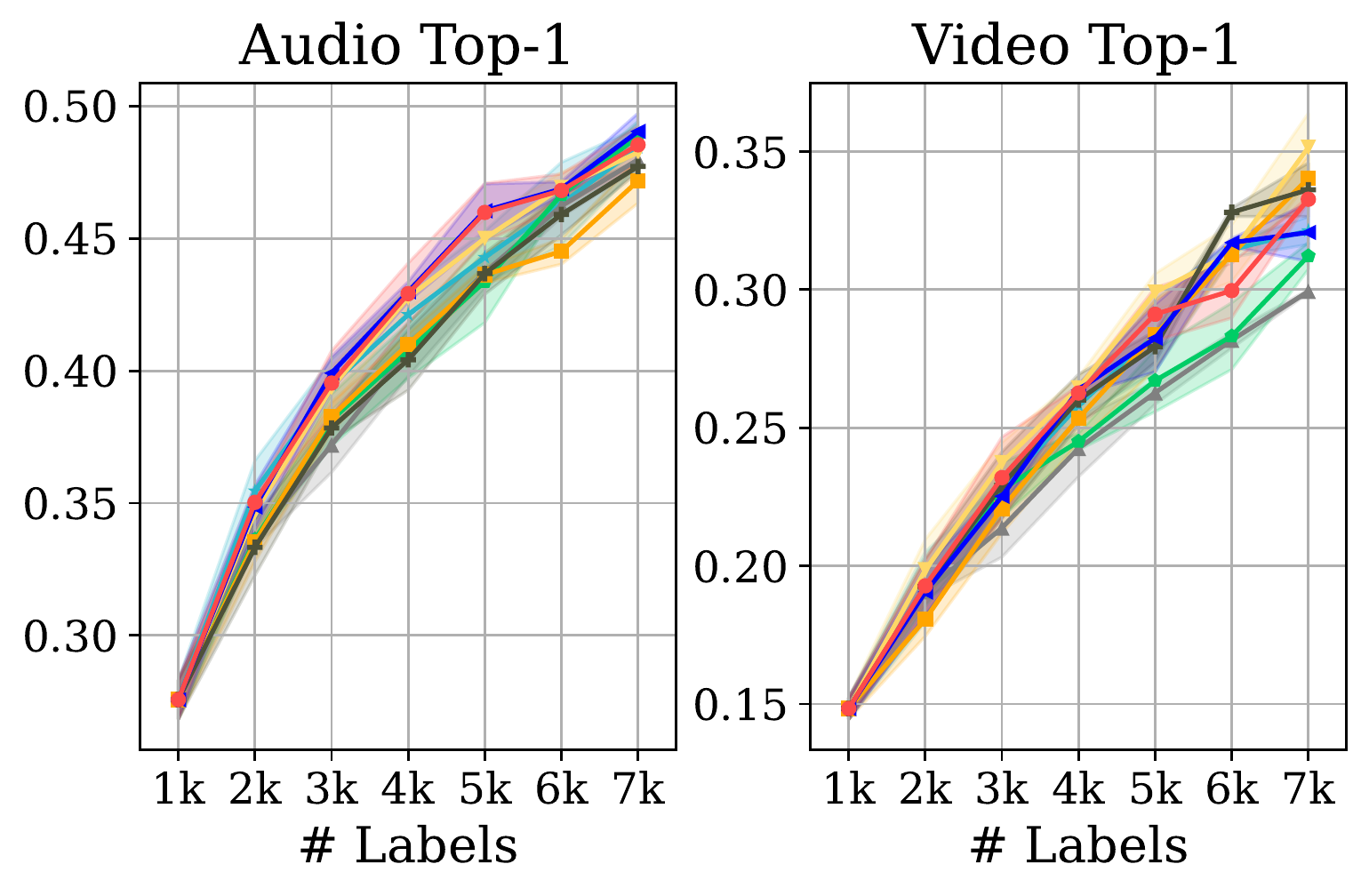}
   \caption{Unimodal performance comparison across AL iterations on KineticsSound.}
   \label{fig:ks-percentage-Top-1-uni}
\end{subfigure}
\caption{Performance comparison between proposed method and other conventional AL strategies with Summation fusion method. The metric selected is top-1 accuracy (Top-1) on mulitmodal and unimodal classification.}
\label{fig:image-text-classification-percentage-Top-1-summation}
\end{figure*}

\begin{figure*}
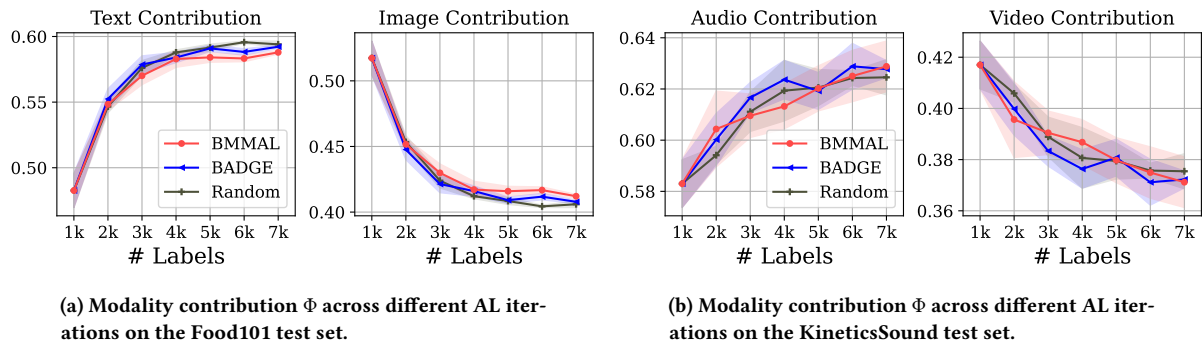

\centering
\begin{subfigure}{.45\textwidth}
  \centering\captionsetup{width=.8\linewidth}
  \includegraphics[width=1.0\linewidth]{figures/food101-contribution.pdf}
  \caption{Modality contribution $\Phi$ across different AL iterations on the Food101 test set.}
  \label{fig:food101-contribution-summation}
\end{subfigure}
\begin{subfigure}{.45\textwidth}
  \centering\captionsetup{width=.8\linewidth}
  \includegraphics[width=1.0\linewidth]{figures/kinetics_sound_2m-contribution.pdf}
  \caption{Modality contribution $\Phi$ across different AL iterations on the KineticsSound test set.}
  \label{fig:kinetics_sound_2m-contribution-summation}
\end{subfigure}
\caption{Unimodal contribution comparison among proposed method, BADGE and random selection with Summation fusion.}
\label{fig:unimodal-contribution-summation}
\end{figure*}

We visualize the performance comparison of all baselines with our proposed method in all AL rounds on Food101 and KineticsSound with fusion mechanism of summation in \textbf{Figure \ref{fig:image-text-classification-percentage-Top-1-summation}}, and the unimodal contribution among BMMAL, BADGE and Random in \textbf{Figure \ref{fig:unimodal-contribution-summation}}. As shown in the figures, our proposed method outperforms BADGE on Food101 and achieves more balanced unimodal contribution than BADGE. While on KineticsSound, our proposed method is comparable with BADGE, and it may be due to the weak fusion ability of summation.

\end{document}